\documentclass[aps,prb,10pt,twocolumn,floatfix,showpacs,superscriptaddress]{revtex4-2}
\usepackage{graphicx}
\usepackage{bbm}
\usepackage{amsmath}
\usepackage{dsfont}
\usepackage{amsfonts}
\usepackage{amsbsy}
\usepackage{physics}
\usepackage{amssymb}
\usepackage{color}
\usepackage{soul}
\usepackage[normalem]{ulem}
\usepackage{placeins} 

\usepackage{caption}
\usepackage{subcaption}
\usepackage[urlcolor=blue,colorlinks=true,citecolor=blue,linkcolor=blue]{hyperref}
\usepackage[nameinlink]{cleveref}
\crefmultiformat{figure}{~#2#1#3}{,\!~#2#1#3}{, #2#1#3}{ and~#2#1#3}
\crefname{equation}{Eq.}{Eqs.}  % Singular and plural forms
\crefname{figure}{Fig.}{Figs.} 
\crefname{section}{Sec.}{Secs.} 
\crefname{appendix}{Appendix}{Appendices}

\usepackage{orcidlink}
\newcommand{\orcidauthorA}{%
  \protect\orcidlink{0000-0002-1207-0407}%
}
\newcommand{\orcidauthorB}{%
  \protect\orcidlink{0000-0003-1400-2165}%
}
\newcommand{\orcidauthorC}{%
  \protect\orcidlink{0000-0001-5207-7235}%
}

\AtBeginDocument{%
	\newwrite\bibnotes
	\def\bibnotesext{Notes.bib}
	\immediate\openout\bibnotes=\jobname\bibnotesext
	\immediate\write\bibnotes{@CONTROL{REVTEX41Control}}
	\immediate\write\bibnotes{@CONTROL{%
			apsrev41Control,author="08",editor="1",pages="1",title="1",year="1"}}
	\if@filesw
	\immediate\write\@auxout{\string\citation{apsrev41Control}}%
	\fi
}%

\begin{document}
%\title{Nonreciprocal supercurrent from many-body branch competition in a parallel double-quantum-dot Josephson junction}
\title{Flux- and tunnel-parity-controlled Josephson nonreciprocity from singlet–triplet competition in a parallel double quantum dot}
\author{Debika Debnath\orcidauthorA}
\email{debika.debnath@tu-braunschweig.de}
\affiliation{Institut für Mathematische Physik, Technische Universität Braunschweig, D-38106 Braunschweig, Germany}

\author{Fernando Dominguez\orcidauthorB}%\orcidB{}
\email{fernando.dominguez@uni-wuerzburg.de}
\affiliation{Institute for Theoretical Physics and Astrophysics, and W\"urzburg-Dresden Cluster of Excellence on Complexity, Topology and Dynamics in Quantum Matter ctd.qmat, Julius-Maximilians-Universit\"at W\"urzburg, Am Hubland, D-97074 W\"urzburg, Germany}

\author{Patrik Recher\orcidauthorC}%\orcidC{}
\email{p.recher@tu-braunschweig.de}
\affiliation{Institut für Mathematische Physik, Technische Universität Braunschweig, D-38106 Braunschweig, Germany}
\affiliation{Laboratory for Emerging Nanometrology Braunschweig, D-38106 Braunschweig, Germany}

\begin{abstract}
{Nonreciprocal supercurrents in interacting nanostructures can arise when the positive and negative critical currents probe inequivalent portions of a phase-dependent many-body spectrum. We study this mechanism in a parallel double-quantum-dot Josephson junction with strong intradot Coulomb repulsion, where local and non-local Cooper-pair transfer coexist. Using exact diagonalization of a zero-bandwidth Hamiltonian with explicit superconducting orbitals, we show that orbital flux and the gauge-invariant tunnel-sign parity control the interference between local Josephson processes and non-local exchange. For detuned dots, this produces flux-tunable singlet, triplet, and doublet ground-state spectral branches and two rectification regimes. Near the boundaries of the charge sector with singly occupied dots [$(1,1)$ sector], doublet branches compete with singlet or triplet branches, yielding diode efficiencies approaching $40\%$. Within the triplet-dominated $(1,1)$ sector, nearby singlet crossings asymmetrically reshape the ground-state envelope even when both critical-current extrema lie on the triplet-like branch, producing a broader response of order $10$--$15\%$. Finite temperature generally suppresses charge-crossover rectification, however, for intermediate temperatures we find an enhancement of both the normalized efficiency and the absolute critical-current asymmetry in the triplet regime. These results establish phase-dependent many-body branch competition as a gate-, flux-, and tunnel-parity-sensitive source of Josephson nonreciprocity.
}
\end{abstract}

\maketitle

\section{Introduction}\label{intro}

The superconducting diode effect is characterized by unequal critical-current magnitudes for opposite transport directions,
$I_\text{c}^+\neq I_\text{c}^-$, creating a polarity-dependent interval of zero-voltage current flow. Following its observation in noncentrosymmetric superconducting films~\cite{Ando2020}, nonreciprocal critical currents have been demonstrated in a variety of Josephson weak links~\cite{Baumgartner2022,Pal2022,Wu2022,Legg2022,Narita2022,Bauriedl2022,Lu2023,He2023,Costa2023,Debnath2024,Yerin2024,Roy2025,Chakraborty2025,Yerin2025,Takeuchi2025,Debnath2025, Debnath2026}. Here, 
broken time-reversal and spatial symmetries allow for a nonreciprocal response, but are not by themselves sufficient~\cite{Nadeem2023, Chen2024, Tatara2026}. Indeed, an anomalous yet purely sinusoidal relation,
\begin{align}
I(\phi)=I_1\sin(\phi -\phi_0)
\end{align}
still has extrema $I_\text{c}^+=I_\text{c}^-=I_1$. Therefore, finite rectification requires an asymmetric current-phase relation, generated, for example, by the interplay of an anomalous phase shift with higher harmonics~\cite{Souto2022, Chen2024}, finite-momentum pairing~\cite{Davydova2022, Fu2022, Yanase2022, Chakraborty2025}, or phase-dependent rearrangements of the Andreev spectrum~\cite{Pillet2023, Mondal2025}.

Superconducting interferometers offer a particularly transparent way to engineer this asymmetry~\cite{Souto2022}. In a two-arm interferometer, magnetic flux ($\Phi$) introduces a relative phase offset between the currents carried by the two arms. If the arms exhibit different harmonic content, the resulting total current
\begin{align}
I(\phi)=I_1\sin(\phi)+I_2\sin(2\phi+\pi\Phi/\Phi_0)
\end{align}
obtains different positive and negative extrema because the phase shift $\delta=\pi\Phi/\Phi_0$ cannot be removed by a change of $\phi$~\cite{Souto2022}. This mechanism was proposed for mesoscopic Josephson interferometers~\cite{Souto2022} and subsequently realized in gate-tunable semiconductor SQUIDs~\cite{Ciaccia2023, Greco2024, Nikodem2025}. In addition to SQUID geometries, multiterminal Josephson junctions extend this principle by providing additional superconducting phases and bias currents as independent control parameters~\cite{Pankratova2020,Matsuo2023, HuamaniCorrea2024,Thieme2026}. 

Quantum-dot Josephson junctions add a qualitatively different ingredient to this interferometric picture by introducing strong Coulomb correlations~\cite{Deacon2015}. Their subgap spectrum can be tuned by gate voltages across charge- and spin-sector transitions, producing highly anharmonic current-phase relations and abrupt changes of the many-body ground state~\cite{Vecino2003a,Grove2018,Debbarma2023}. A parallel double quantum dot is especially rich because a Cooper pair may be transferred locally through one dot or split between the two arms. When both dots are singly occupied, the corresponding non-local virtual processes generate an effective exchange interaction between the two dot spins, allowing singlet and triplet ground states whose ordering depends on the superconducting phase, magnetic flux, and the relative signs of the tunnel amplitudes~\cite{Choi2000,Wang2011,Probst2016}. 

Recent work on the same parallel-double-dot geometry identified a branch-selected Josephson diode effect across a singlet--doublet $0$--$\pi$ boundary and established the importance of non-local Cooper-pair transfer in the weakly interacting limit $U/\Delta\ll 1$~\cite{Zhang2026}. Here, we consider the strongly interacting regime $U/\Delta\gg 1$ and investigate the complementary role of exchange-driven singlet--triplet competition mediated by the coupling to the superconductors. Using a zero-bandwidth formulation of the superconducting leads solved by exact diagonalization and supplemented by a fourth-order analysis of the tunneling processes, we determine the phase-dependent many-body spectrum and the associated Josephson current. We show that the orbital flux and the gauge-invariant tunnel parity $\mathcal{P}$ jointly control the ordering of the singlet and triplet sectors: the two local Josephson channels acquire opposite flux-induced phase shifts, whereas non-local cotunneling modifies the exchange splitting. The resulting competition between singlet, triplet, and doublet branches determines the equilibrium ground-state envelope and produces a nonreciprocal critical current. 

We identify two distinct finite-flux rectification regimes. Near the boundaries of the sector with a single electron on each dot, doublet branches whose phase dispersion is dominated by local Josephson processes compete with singlet or triplet branches receiving substantial non-local pair-transfer and exchange contributions. Here, the phase-dependent ground state branch selection distorts strongly the current--phase relation and yields diode efficiencies approaching $\eta=40\%$. A similar regime was observed also in the weakly interacting regime~\cite{Zhang2026}. Additionally, we find a new regime of broken reciprocity extending broadly within the triplet-dominated $(1,1)$ sector. There, nearby singlet crossings delimit the phase interval over which the triplet-like branch forms the equilibrium ground state and reshape its positive and negative current extrema differently, yielding efficiencies of order $\eta=15\%$ at zero temperature. Remarkably, the latter case is only present in the strongly interacting regime, i.e.~$U\gg \Delta$.

At finite temperature, thermal occupation redistributes the contributions of Andreev states to the supercurrent and can substantially modify, or even enhance, the Josephson diode efficiency~\cite{Kokkeler2022, Maiani2023, Ghosh2024, Yang2025}. For the parameters studied here, the diode efficiency $|\eta|$ decreases monotonically near charge-sector crossovers, whereas it exhibits a nonmonotonic response with intermediate thermal enhancement in the triplet-dominated regime. We interpret these trends through the energy-current covariance at the global critical phase differences, which describes how thermal redistribution among competing many-body states modifies the relative imbalance between the critical currents. Enhancement or suppression therefore depends on the excitation spectrum and the currents carried by its states, and cannot be inferred solely from the spin character of the ground state.

\begin{figure}[t]
    \centering
\includegraphics[width=3.3 in%width=0.99\linewidth,height=0.21\textwidth
]{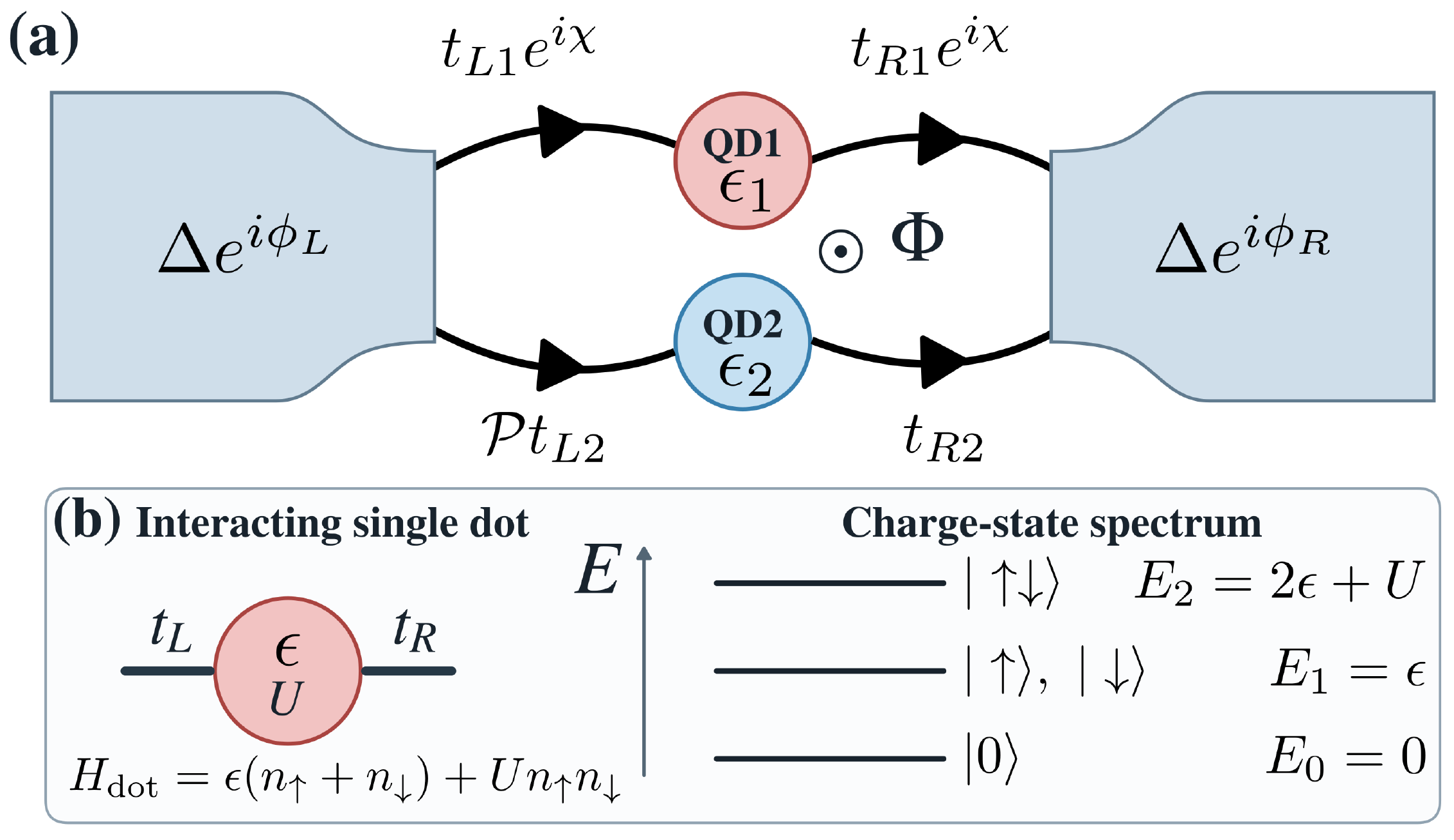}
\caption{Parallel double-quantum-dot Josephson junction and single-dot charge states. (a) Two quantum dots with level energies $\epsilon_i$ where $i (=1,2)$ and charging energies $U_i$ are coupled in parallel between superconducting reservoirs with phases $\phi_L$ and $\phi_R$. The magnetic flux $\Phi$ threading the loop generates the single-electron Aharonov--Bohm phase $\Phi$. The four lead-dot couplings $t_{\nu i}$ define the gauge-invariant tunnel parity $\mathcal{P}$, introduced in Eq.~\eqref{eq:tunnel-parity} and $\chi=\pi\Phi/2\Phi_0$. (b) Interacting single-level quantum dot described by $H_{\mathrm{dot}}$ with empty, singly occupied, and doubly occupied states of energies $E_0=0$, $E_1=\epsilon$, and $E_2=2\epsilon+U$, respectively.}
    \label{fig:1}
\end{figure}

The remainder of the paper is organized as follows. In Sec.~\ref{Model&Formalism}, we introduce the microscopic model, the zero-bandwidth formulation, the definition of the diode response, and the relevant charge and spin sectors. 
Sec.~\ref{sec:results} presents the results for the flux- and tunnel-parity-controlled interference mechanism, gate-controlled diode response and finite temperature effects. In Sec.~\ref{sec:conclusions}, we summarize our findings. Details of the calculations and interpretations as well as the effects of a Zeeman splitting on the dots and spin-orbit coupling are discussed in several appendices.

\section{Hamiltonian, zero band width approximation and superconducting diode effect} 
\label{Model&Formalism}

\subsection{Model Hamiltonian}
\label{sec:model-hamiltonian}

We consider two interacting, single-level QDs connected in parallel between two conventional $s$-wave superconductors, as sketched in Fig.~\ref{fig:1}. The Hamiltonian is given by
\begin{equation}
    H=H_{S}+H_{\mathrm{DQD}}+H_{T}.
    \label{eq:H}
\end{equation}

The two BCS reservoirs are described by
\begin{equation}
\begin{aligned}
    H_{S}=\sum_{\nu=L,R}\Bigg[&\sum_{\mathbf{k}\sigma}
        \xi_{\mathbf{k}\nu}
        c_{\mathbf{k}\nu\sigma}^{\dagger}c_{\mathbf{k}\nu\sigma}\\
        &+\sum_{\mathbf{k}}
        \left(
        \Delta e^{-i\phi_{\nu}}
        c_{\mathbf{k}\nu\uparrow}^{\dagger}
        c_{-\mathbf{k}\nu\downarrow}^{\dagger}
        +\mathrm{h.c.}\right)\Bigg],
\end{aligned}
\label{eq:H_lead}
\end{equation}
with $\nu=L,R$ and single-particle (momentum $\mathbf{k}$ and spin $\sigma$) energy $\xi_{\mathbf{k}\nu}$ is measured from the common chemical potential. Here, the two leads have the same pairing energy $\Delta$ and phases $\phi_L$ and $\phi_R$. We denote their bare phase difference by $\phi=\phi_L-\phi_R$.

Each dot is represented by a spin-degenerate Anderson orbital
\begin{equation}
    H_{\mathrm{DQD}}
    =\sum_{i=1,2}\left[
        \epsilon_i \hat N_i
        +U_i \hat n_{i\uparrow}\hat n_{i\downarrow}
    \right],
    \label{eq:H_QD}
\end{equation}
with $\hat n_{i\sigma}=d_{i\sigma}^{\dagger}d_{i\sigma}$ and $\hat N_i=\hat n_{i\uparrow}+\hat n_{i\downarrow}$. The level positions $\epsilon_i$ are gate tunable and $U_i>0$ is the intradot repulsive Coulomb energy. The corresponding charge sectors and spin multiplets are summarized in Sec.~\ref{sec:charge-spin}.

In this model, we have neglected interdot coupling terms, such as tunneling $\tau_{12}d^\dagger_{1\sigma}d_{2\sigma}$ or the 
Coulomb coupling $U_{12} \hat{n}_{1 \sigma}\hat{n}_{2 \sigma'}$. The presence of these terms can modify the results presented in this contribution quantitatively, but the general behavior will still be present. 

The tunneling Hamiltonian between the QDs and the superconducting leads is given by
\begin{equation}
    H_T=\sum_{\nu\mathbf{k}i\sigma}
    \left(
        \tau_{\nu i}
        d_{i\sigma}^{\dagger}c_{\mathbf{k}\nu\sigma}
        +\mathrm{h.c.}
    \right),
    \label{eq:H_T}
\end{equation}
with the tunneling amplitudes $\tau_{\nu i}$. Here, the orbital magnetic field enters through the Peierls substitution in the tunneling amplitudes. Denoting by $\mathbf r_i$ the position of dot $i$ and by $\mathbf r_{\nu i}$ the corresponding contact point on lead $\nu$, we write
\begin{equation}
\tau_{\nu i}
=
t_{\nu i}^{(0)}e^{i a_{\nu i}},
\qquad
a_{\nu i}
=
\frac{\pi}{\Phi_0}
\int_{\mathbf r_i}^{\mathbf r_{\nu i}}
d\mathbf l\cdot\mathbf A(\mathbf l),
\label{eq:peierls-phase-model}
\end{equation}
where $t_{\nu i}^{(0)}$ is the real-valued zero-field tunneling amplitude and $\Phi_0=h/2e$. The integration path follows the corresponding dot--lead tunnel link. Local rephasings of the lead and dot operators can redistribute the phases and the signs among the four tunneling amplitudes. The remaining gauge-invariant sign is their product
\begin{equation}
\mathcal P
\equiv
\operatorname{sgn}
\left[
t_{L1}^{(0)}
t_{R1}^{(0)}
t_{L2}^{(0)}
t_{R2}^{(0)}
\right]
=\pm1,
\label{eq:tunnel-parity}
\end{equation}
which we refer to as the \emph{tunnel parity}. Note that the physical origin of the signs in $t_{\nu i}^{(0)}$ arises from the relative sign between the overlapping quantum dot and superconductor wave functions. Indeed, this sign can change from dot level to dot level and therefore it can be tuned by means of a gate voltage, see further details in Ref.~\onlinecite{Probst2016}. In the remaining of this contribution, we choose a gauge in which this sign is assigned to the $L2$ link and the Peierls phases are placed on the two links connected to dot 1, that is
\begin{align}
\tau_{L1}&=t_{L1}e^{i\chi},~~~
\tau_{R1}=t_{R1}e^{-i\chi},\label{eq:tunnel-amplitudes-model_1}\\
\tau_{L2}&=\mathcal P t_{L2},~~~
\tau_{R2}=t_{R2},
\label{eq:tunnel-amplitudes-model}
\end{align}
with $\chi\equiv\frac{\pi\Phi}{2\Phi_0}$ and $t_{\nu i}=|t_{\nu i}^{(0)}|$. In this gauge, the single-electron Aharonov--Bohm phase accumulated around the loop (clockwise)  is
\begin{equation}
\left(a_{L1}-a_{R1}\right)
-
\left(a_{L2}-a_{R2}\right)
=
2\chi
=
\frac{\pi\Phi}{\Phi_0}.
\end{equation}

Following Ref.~\cite{Choi2000}, we introduce the gauge-invariant superconducting phase drop across each of the two arms. For the path $\mathcal C_i$ running from the right reservoir to the left reservoir through dot $i$, these phases are
\begin{align}
\varphi_i
&=
\phi_L(\mathbf r_{Li})
-
\phi_R(\mathbf r_{Ri})
-
\frac{2\pi}{\Phi_0}
\int_{\mathcal C_i}
d\mathbf l\cdot\mathbf A(\mathbf l)
\
\\
&=
\phi_L-\phi_R
-
2\left(a_{Li}-a_{Ri}\right),
\qquad i=1,2,
\label{eq:arm-gauge-phases-model}
\end{align}
where the second line follows for reservoirs with spatially uniform superconducting phases. The two phase drops are individually gauge invariant but are constrained by the enclosed flux. With the orientation adopted above
\begin{equation}
\varphi_2-\varphi_1
=
\frac{2\pi\Phi}{\Phi_0}.
\label{eq:arm-phase-constraint}
\end{equation}
For the gauge in Eqs.~\eqref{eq:tunnel-amplitudes-model_1}-\eqref{eq:tunnel-amplitudes-model}, they reduce to
\begin{equation}
\varphi_1
=
\phi_L-\phi_R-2\pi\frac{\Phi}{\Phi_0},
\qquad
\varphi_2
=
\phi_L-\phi_R.
\end{equation}
It is convenient to parameterize the current--phase relation by their mean
\begin{equation}
\begin{aligned}
\bar{\varphi}
&\equiv
\frac{\varphi_1+\varphi_2}{2}
=
\phi_L-\phi_R-\frac{\pi\Phi}{\Phi_0},~~\text{and the flux}~~\frac{\pi\Phi}{\Phi_0}, 
\label{eq:gauge-phase-model}
\end{aligned}
\end{equation}
resulting 
\begin{align}
\varphi_1
=
\bar{\varphi}-\frac{\pi\Phi}{\Phi_0},
\qquad
\varphi_2
=
\bar{\varphi}+\frac{\pi\Phi}{\Phi_0}.
\end{align}
In this gauge, local Cooper-pair transfer through dot $i$ consequently depends on the corresponding arm phase $\varphi_i$, whereas non-local split-pair transfer depends on the mean phase $\bar{\varphi}$. All current--phase relations below are therefore expressed as functions of $\bar{\varphi}$ at fixed orbital flux. For simplicity, we remove from now on the bar, $\bar{\varphi}\rightarrow \varphi$.

This model neglects additional geometrical suppression of non-local transfer. Therefore, its validity restricts to the limit where the DQD are separated by a distance smaller than the superconducting coherence length $\xi_s=\hbar v_F/\Delta$, with $v_F$ the Fermi velocity of the superconducting leads~\cite{Choi2000, Recher2001, Recher2002}. 
 
\subsection{Zero-bandwidth approximation}
\label{sec:zba}

In order to retain the interacting subgap spectrum while keeping the Hilbert space finite, we compose an effective low energy model by means of the zero-bandwidth approximation (ZBA)~\cite{Affleck2000,Probst2016,Grove2018,Debbarma2023,Yi2024,Zalom2024}.  In this approximation, each superconducting continuum is replaced by one spinful orbital at the Fermi energy with an on-site pairing term
\begin{equation}
    H_S^{\mathrm{ZBA}}
    =\sum_{\nu=L,R}
    \left(
        \Delta_b e^{-i\phi_\nu}
        c_{\nu\uparrow}^{\dagger}c_{\nu\downarrow}^{\dagger}
        +\mathrm{H.c.}
    \right).
    \label{eq:H-zba-leads}
\end{equation}
Thus, the tunneling Hamiltonians retain the phase and spin structure of Eqs.~\eqref{eq:H_T}, with $c_{\mathbf{k}\nu\sigma}\rightarrow \delta_{\mathbf{k},0}c_{\nu\sigma}$ and renormalized parameters $\Delta\rightarrow \Delta_b$ and $t_{\nu i}\rightarrow\bar t_{\nu i}$. When matching the effective model to a flat-band continuum with normal-state density of states $\rho_0$ per spin, one would need to use the hybridization scale
\begin{equation}
   \Gamma_{\nu i}=\pi\rho_0\abs{\bar t_{\nu i}}^2,
   \label{eq:hybridization}
\end{equation}
and $\Gamma_i=\sum_\nu\Gamma_{\nu i}$. Moreover, the renormalized parameters $\Delta_b$ and $\bar{t}_{\nu i}$ may be obtained from self-consistent calculations~\cite{Affleck2000}.

The ZBA treats the Coulomb interactions and all dot charge sectors exactly, retaining the discrete many-body states generated by superconducting pairing and lead-mediated exchange~\cite{Vecino2003a, Bergeret2007a}. Importantly, the singlet–triplet transition relevant to the present work is not specific to the minimal zero-bandwidth approximation. For the same parallel-DQD Josephson geometry, Ref.~\onlinecite{Probst2016} compared exact-diagonalization results within the ZBA with fourth-order tunneling calculations for continuum, wide-band BCS reservoirs. The triplet ground-state region was obtained in both descriptions, without qualitative differences in the phase diagram, and was traced to triplet-
favoring nonlocal cotunneling processes at finite charging energy. We therefore use the minimal ZBA to resolve the complete phase-dependent many-body spectrum and its finite-temperature occupation. Continuum effects may nevertheless shift the transition lines and current amplitudes quantitatively.

The effective Hamiltonian resulting from the ZBA contains four spinful orbitals (two dots and two superconducting sites) or equivalently, eight fermionic modes. Its Fock space therefore has dimension $2^8=256$. Particle number is not conserved because of the pairing terms, but total fermion parity is. Throughout this work, we consider the equilibrium Josephson response and select the global many-body ground state at each superconducting phase. This construction assumes that quasiparticle and spin relaxation are sufficiently fast for the junction to switch between competing parity and spin sectors. For a phase sweep performed at fixed fermion parity or in the absence of spin relaxation, the system would instead follow a constrained or metastable branch, leading in general to a different current-phase relation.

At zero temperature, the equilibrium Josephson current follows from the lowest many-body energy
\begin{equation}
    I(\varphi)
    =\frac{2e}{\hbar}
    \left.\frac{\partial E_{\mathrm{GS}}}{\partial\varphi}\right|_{\Phi},
    \label{eq:I}
\end{equation}
with the ground state energy $E_{\mathrm{GS}}$. At a crossing, $E_{\mathrm{GS}}$ is the lower envelope of the competing branches and the derivative is taken branchwise, so that the current can be discontinuous.

At finite temperature, $E_{\mathrm{GS}}$ is replaced by the free energy, i.e.~$ I(\varphi)
    =\frac{2e}{\hbar}
    \partial F/\partial\varphi|_{\Phi}$ with $F=-k_B T \ln Z(\varphi,T)$, and the partition function $Z=\sum_n \exp(-\beta E_n)$ and $\beta=1/k_BT$. In this way, we can express the finite temperature supercurrent as
\begin{align}\label{eq:curr_finiteT}
   I(\varphi)
    =\frac{2e}{\hbar}\sum_n p_n
    \langle n| \frac{\partial H}{\partial\varphi}|n\rangle,
\end{align}
with $p_n= \exp(-\beta E_n)/Z$. 

In the rest of this contribution, we show results of the dimensionless supercurrent $\mathcal{I}=\hbar I/(2e\Delta_b)$ and critical currents $\mathcal{I}_\text{c}$ and $\mathcal{I}^{\pm}_\text{c}$. In the numerical results, energies are measured in units of the effective gap, $\Delta_b$, and the bars on the effective tunnel parameters are omitted. Unless stated otherwise, we set $\epsilon_2=-1.5\Delta_b$, $U_1=U_2\equiv U$, and $t_{L1}=|t_{L2}|=t_{R1}=t_{R2}=0.5\Delta_b$. 

\subsection{Josephson diode response}
\label{sec:diode-definition}
%A Josephson diode has unequal maximum supercurrents in the two transport directions, i.e.~$\mathcal{I}^{+}_\text{c}\neq \mathcal{I}^{-}_\text{c}$. 
In the present double-dot interferometer, orbital flux breaks time-reversal symmetry, while unequal dot energies or tunnel amplitudes break the symmetry that exchanges the two arms. 

We define
$\mathcal{I}_c^+=\max_{\varphi}\mathcal{I}(\varphi)$ and
$\mathcal{I}_c^-=-\min_{\varphi}\mathcal{I}(\varphi)$
as positive critical-current magnitudes and quantify the diode response by
\begin{equation}
    \eta=
    \frac{\mathcal{I}_c^+-\mathcal{I}_c^-}{\mathcal{I}_c^++\mathcal{I}_c^-}\times100\%.
    \label{eq:effi}
\end{equation}
Thus, $\eta=0$ denotes reciprocal transport, while the sign of $\eta$ identifies the favored current direction.

\subsection{Relevant energy scales and many-body sectors}
\label{sec:charge-spin}

Here, we describe some general aspects concerning the occupation number, total spin and Cooper pair transfer paths taking place in the DQD Josephson junction. 

\textbf{Charge sectors:}
Each QD is modeled by a single spin-degenerate Anderson orbital. We assume that the nearest neglected orbital lies outside the energy window of interest and is separated by $\delta\epsilon_i$. With $\Gamma_i=\sum_{\nu=L,R}\Gamma_{\nu i}$ the total tunnel broadening of dot $i$, the single-orbital Coulomb-blockade description requires $\Gamma_i\ll\delta\epsilon_i$ and $\Gamma_i\ll U_i$. Finite tunnel coupling nevertheless admixes neighboring charge configurations. Accordingly, $(N_1,N_2)$ denotes the dominant DQD charge sector. These labels are exact well inside a Coulomb-blockade valley and should be understood as the dominant character close to a charge or spin transition, where hybridization mixes different configurations.

Taking the common chemical potential of the superconducting leads as the zero of energy, the electrostatic energy of an isolated dot containing $N_i=0,1,2$ electrons is
\begin{equation}
    \mathcal{E}_i(N_i)
    =N_i\epsilon_i+\frac{U}{2}N_i(N_i-1),
    \label{eq:atomic-dot-energy}
\end{equation}
i.e., $\mathcal{E}_i(0)=0$, $\mathcal{E}_i(1)=\epsilon_i$, and $\mathcal{E}_i(2)=2\epsilon_i+U$. Minimization of Eq.~\eqref{eq:atomic-dot-energy} gives the atomic-limit occupation
\begin{equation}
    N_i^{\rm at}=\begin{cases}
        0, & \epsilon_i>0,\\
        1, & -U<\epsilon_i<0,\\
        2, & \epsilon_i<-U.
    \end{cases}
    \label{eq:atomic-occupation}
\end{equation}
Thus, the boundaries $\epsilon_i=0$ and $\epsilon_i=-U$ correspond to the $0\leftrightarrow1$ and $1\leftrightarrow2$ charge degeneracies, respectively, while $\epsilon_i=-U/2$ is the particle-hole-symmetric point of the singly occupied quantum dot. Hybridization shifts and broadens these atomic boundaries on the scale of $\Gamma_i$. For the parameter scans used below, $\epsilon_2=-1.5\Delta_b$ and $U_1=U_2=U$, hence $N_2=1$ requires $U>1.5\Delta_b$, whereas for $0<U<1.5\Delta_b$ the second dot is doubly occupied. For example, whenever $N_2=1$, increasing $\epsilon_1$ drives the sequence $(2,1)\rightarrow(1,1)\rightarrow(0,1)$ across $\epsilon_1=-U$ and $\epsilon_1=0$.
 
\textbf{ Spin multiplets:}
An empty or doubly occupied dot carries no spin. Charge sectors with one unpaired dot electron, such as $(0, 1)$ or $(2, 1)$, exhibit spin 1/2  and therefore form doublets. In the $(1, 1)$ sector, the two dot spins combine into a singlet and a triplet manifold
\begin{equation}
\begin{aligned}
    \ket{S}&=\frac{1}{\sqrt{2}}
        \left(\ket{\uparrow,\downarrow}-\ket{\downarrow,\uparrow}\right),\\
    \ket{T_0}&=\frac{1}{\sqrt{2}}
        \left(\ket{\uparrow,\downarrow}+\ket{\downarrow,\uparrow}\right),\\
    \ket{T_+}&=\ket{\uparrow,\uparrow},\qquad
    \ket{T_-}=\ket{\downarrow,\downarrow}.
\end{aligned}
\label{eq:singlet-triplet-basis}
\end{equation}
These four states are degenerate in the isolated DQD because $H_{\mathrm{DQD}}$ contains neither direct interdot hopping nor a bare exchange interaction. Virtual tunneling through the superconductors lifts this degeneracy. Using the convention employed below, $E_T-E_S=2J_{\mathrm{ex}}$, so that $J_{\mathrm{ex}}>0$ favors the singlet and $J_{\mathrm{ex}}<0$ favors the triplet~\cite{Probst2016}. The exchange depends on the tunnel amplitudes, their total parity, the superconducting phase, and the orbital flux. 

\textbf{ Local and non-local Cooper pair transfer:} 
In the parallel geometry, Cooper-pair transfer can proceed locally and non-locally. In a local Josephson process, both electrons pass through the same dot, whereas in a non-local process, the two electrons of a Cooper pair pass through different dots. These labels describe tunneling paths rather than distinct eigenstate quantum numbers: both types of processes can renormalize the same many-body branch. In addition to the Josephson processes, there are cotunneling processes involving virtual charge fluctuations without transferring a Cooper pair between the reservoirs. When both dots participate, it contributes to the effective singlet–triplet exchange in the $(1, 1)$ sector. As we will see below, all these processes react differently to the presence of $\varphi$ and $\Phi$. 

\begin{figure}[t]
    \centering
\includegraphics[width=0.98\linewidth,height=0.38\textwidth]{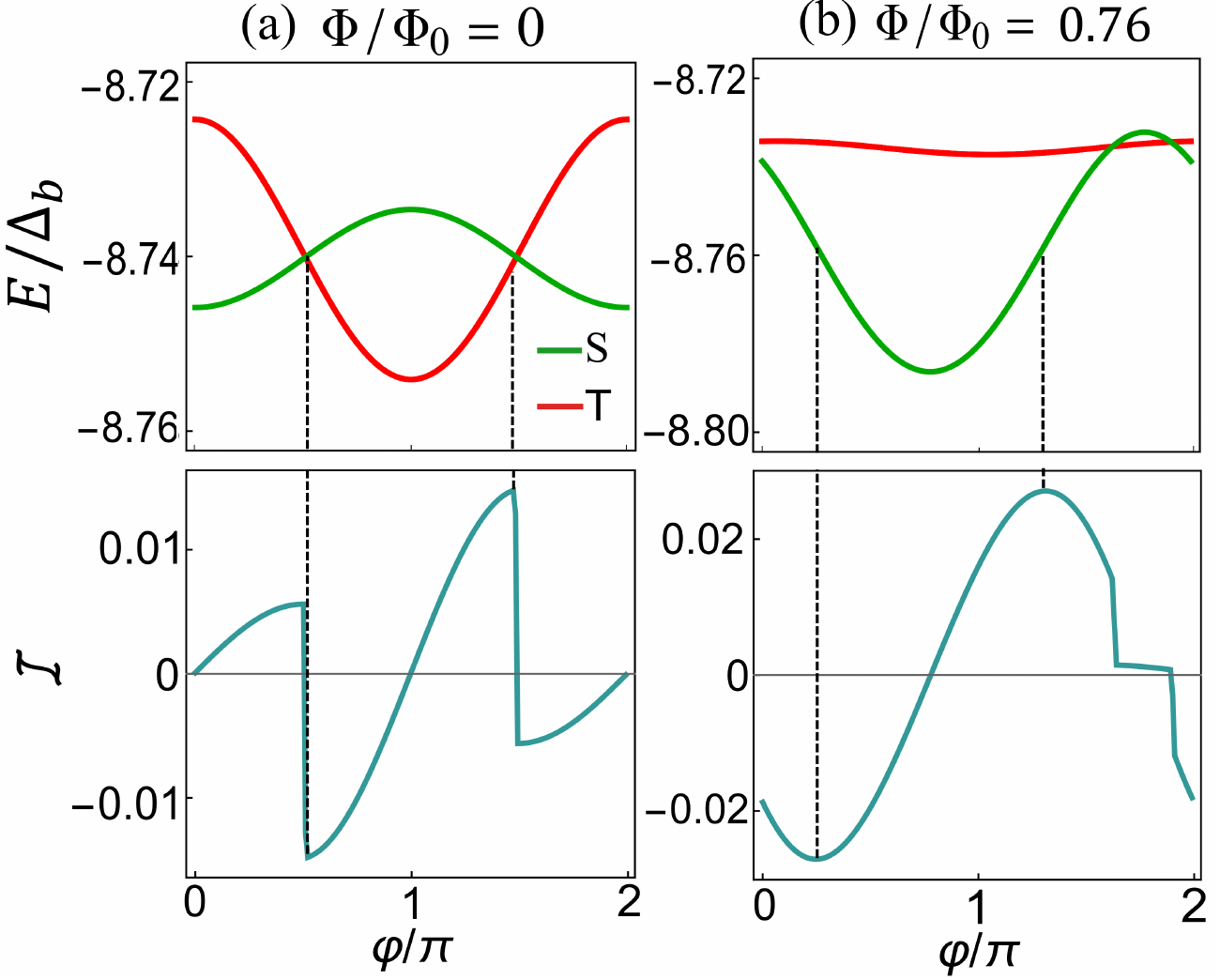}
\caption{Lowest singlet (green) and triplet (red) many-body energies (upper row) and the corresponding Josephson current (lower row), shown as functions of the gauge-invariant phase $\varphi$ at even total tunnel parity $(\mathcal{P}=1)$ for (a) $\Phi/\Phi_0=0$ and (b) $\Phi/\Phi_0=0.76$, respectively. The parameters are $\epsilon_1=-5\Delta_b$, $\epsilon_2=-1.5\Delta_b$, and $U=8\Delta_b$.
}
    \label{fig:2}
\end{figure}

\section{Results}\label{sec:results}
\label{sec:flux-results}

We first determine the gauge-invariant phase and flux dependence of the local and non-local virtual processes and use it to identify the role of the tunnel parity. We then analyze the global flux--gate interference pattern and the microscopic origin of the two rectification regimes, before turning to their distinct finite-temperature responses. Throughout this section, charge and spin labels denote the dominant character of a many-body eigenstate close to a transition and the exact eigenstates contain hybridized charge configurations.

\subsection{Gauge-invariant phase structure and fixed-branch reciprocity}

In the presence of an orbital flux, the total single-electron Aharonov--Bohm phase accumulated around the loop is $2\chi=\pi\Phi/\Phi_0$, and the physical superconducting phase difference is the gauge-invariant phase $\varphi$ defined in Eq.~\eqref{eq:gauge-phase-model}. Expressed in terms of $\varphi$, the fourth-order tunneling contributions have the phase structure
\begin{equation}
\begin{aligned}
    E_J^{\mathrm{nl}}&\propto \mathcal{P}\cos\varphi,
    &E_{\mathrm{CT}}^{\mathrm{nl}}&\propto \mathcal{P}\cos(\pi\Phi/\Phi_0),\\
    E_{J,1}^{\mathrm{loc}}&\propto \cos(\varphi-\pi\Phi/\Phi_0),
    &E_{J,2}^{\mathrm{loc}}&\propto \cos(\varphi+\pi\Phi/\Phi_0), 
\end{aligned}
\label{eq:flux-structure-results}
\end{equation}
whereas the local cotunneling contributions are independent of $\varphi$. The derivation of these structures and the corresponding branch-dependent coefficients is given in App.~\ref{app:gauge_products}. 
%and \ref{app:spin_decomposition}. 
Eq.~\eqref{eq:flux-structure-results} separates three effects. First, at fixed $\varphi$, the non-local (nl) Josephson (J) contribution carries no additional orbital-flux dependence, although it is flux shifted when written in terms of the bare lead-phase difference $\phi_L-\phi_R$~\cite{Choi2000,Wang2011,Haidekker2020a,Dominguez2024a,Zhang2026}. Second, the two local (loc) Josephson channels acquire opposite flux-induced phase shifts. Third, the non-local cotunneling (CT) term retains a genuine Aharonov--Bohm dependence. It does not carry current directly, but it shifts the many-body energies and thereby controls the positions of phase-dependent branch crossings.

In the $(1,1)$ charge sector, the spin-exchange contributions enter the singlet and triplet energies with opposite signs. With the convention used in the perturbative decomposition,
\begin{align}
    &E_T-E_S=2J_{\mathrm{ex}},\qquad \text{and}\nonumber \\
    &J_{\mathrm{ex}}(\varphi,\Phi)=J_0+\mathcal{P}J_J\cos\varphi
    +\mathcal{P}J_{\mathrm{CT}}\cos(\pi\Phi/\Phi_0),
    \label{eq:exchange-results}
\end{align}
so that $J_{\mathrm{ex}}>0$ favors the singlet and $J_{\mathrm{ex}}<0$ favors the triplet. The exchange sign is not fixed by the singlet character of the BCS reservoirs: mixed electron--hole cotunneling sequences can instead stabilize a triplet ground state in the interacting double dot~\cite{Probst2016}, see Apps.~\ref{app:spin_decomposition} and \ref{app:ST_crossing} for further details. Orbital flux and tunnel parity therefore tune the singlet--triplet ordering through the factor $\mathcal{P}\cos(\pi\Phi/\Phi_0)$.

The same decomposition also clarifies what is insufficient for rectification. Up to fourth order, the phase dependence of any many-body branch $a$ can be written as
\begin{equation*}
    E_a^{(4)}(\varphi,\Phi)=C_a(\Phi)+X_a(\Phi)\cos\varphi
    +Y_a(\Phi)\sin\varphi .
\end{equation*}
Its current is therefore a shifted first harmonic and obeys $\mathcal{I}_{c,a}^{+}=\mathcal{I}_{c,a}^{-}$. Thus, orbital flux and arm asymmetry can generate an anomalous phase shift on an individual fourth-order branch, but not a diode response, see fourth-order terms in App.~\ref{app:gauge_products}. At this order, nonreciprocity requires phase-dependent selection among branches that do not share a common reflection center. The exact zero-bandwidth calculation additionally retains higher-order harmonics, which can further skew the current-phase relation.% A detailed derivation is given in App.~\ref{}.

Fig.~\ref{fig:2} illustrates how these ingredients act on the low-energy spectrum. At zero flux, the singlet--triplet crossings occur symmetrically within the phase cycle, and the cusps in the equilibrium current--phase relation do not produce unequal critical-current magnitudes. A finite flux changes both the phase-independent exchange offset and the relative phase of the local Josephson channels. The singlet--triplet crossings and the associated current discontinuities consequently move to inequivalent phases, removing the common reflection center of the ground-state envelope. Reversing the tunnel parity changes the sign of the non-local terms and produces the complementary sequence of singlet- and triplet-dominated phase intervals (not shown).

\subsection{Flux- and tunnel-parity-controlled interference}

\begin{figure}[t]
    \centering
    \includegraphics[width=1.03\linewidth,height=0.45\textwidth]{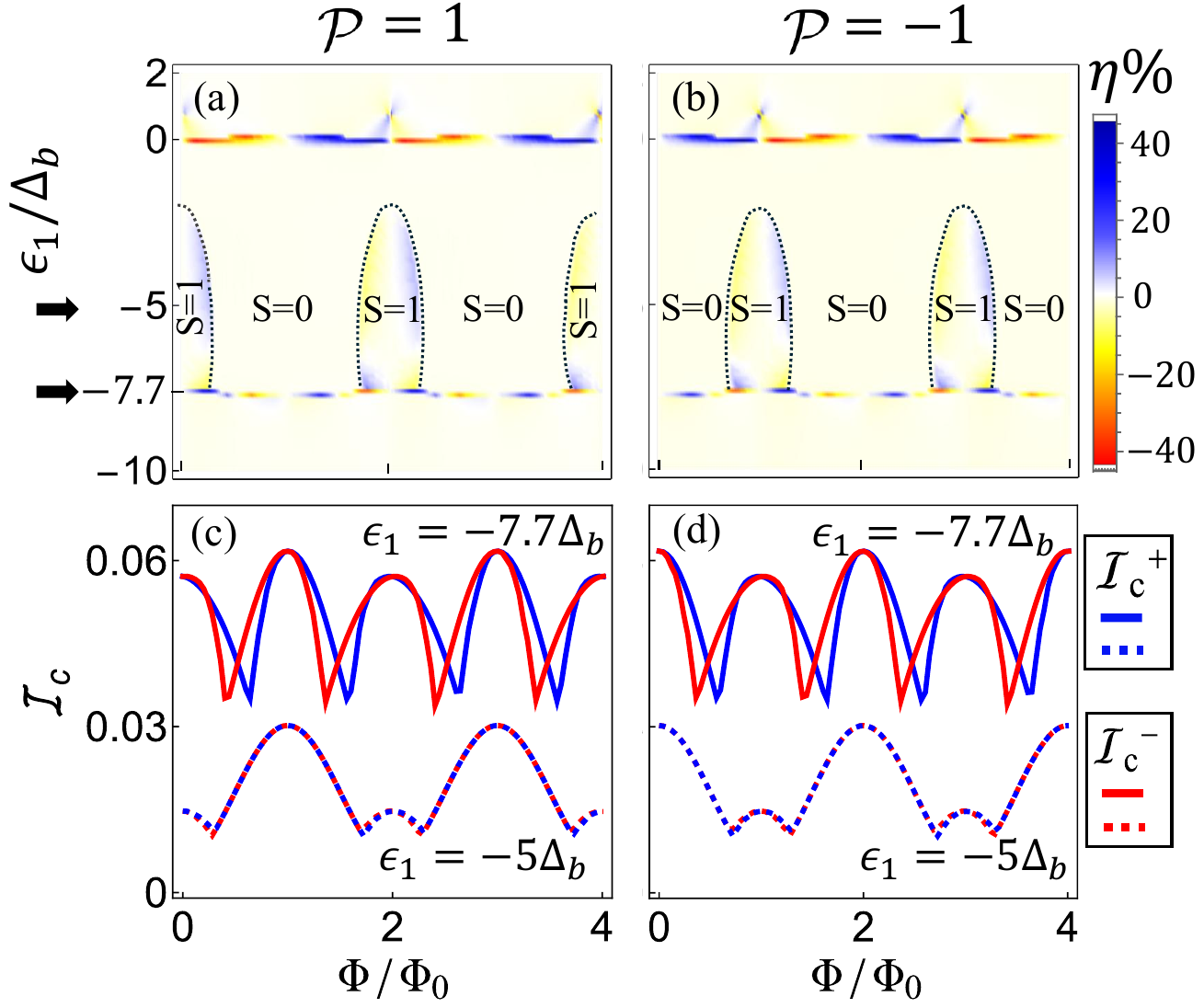}
    \caption{Josephson-diode efficiency as a function of the first-dot energy $\epsilon_1$ and magnetic flux $\Phi/\Phi_0$ for (a) even ($\mathcal{P}=1$) and (b) odd ($\mathcal{P}=-1$) tunnel parity at $U=8\Delta_b$. Dashed curves mark boundaries between singlet- and triplet-dominated ground-state envelopes within the $(1,1)$ charge sector. Panels (c) and (d) show $I_c^+$ (blue) and $I_c^-$ (red) for even and odd tunnel parity, respectively, along the line cuts shown by the arrows at $\epsilon_1=-5\Delta_b$ (dashed) and $\epsilon_1=-7.7\Delta_b$ (solid). 
    %\textcolor{red}{Indicate line cuts in panels a, b}
    }
    \label{fig:3}
\end{figure}

The global consequences of the exchange and interference structure are summarized in Fig.~\ref{fig:3}. The maps combine the flux-dependent spin ordering inside the $(1,1)$ sector with the SQUID-like modulation of the critical currents.

For $\mathcal P=1$, the triplet-dominated regions occur predominantly around even multiples of $\Phi_0$, whereas for $\mathcal P=-1$ they are displaced toward odd multiples of $\Phi_0$, with the singlet occupying the complementary flux intervals. This behavior follows from the non-local cotunneling contribution $\mathcal P\cos(\pi\Phi/\Phi_0)$ in Eq.~\eqref{eq:flux-structure-results}. Indeed, reversing $\mathcal P$ is equivalent in this factor to shifting the flux by one superconducting flux quantum, $\Phi\rightarrow\Phi+\Phi_0$, that is
\begin{equation}
    E_{-\mathcal P}(\varphi,\Phi)
    =
    E_{\mathcal P}(\varphi+\pi,\Phi+\Phi_0).
\end{equation}
The phase-resolved spectra for the two tunnel parities are therefore related by both a half-period flux translation and a rigid $\pi$ shift of the superconducting phase. Since the critical currents and diode efficiency are invariant under such a rigid phase translation, their flux maps satisfy
$\eta_{-\mathcal P}(\Phi)
=\eta_{\mathcal P}(\Phi+\Phi_0)$.

Due to the geometry of the Josephson junction, the critical current as a function of the magnetic flux shown in Figs.~\ref{fig:3}(c) and \ref{fig:3}(d) display a SQUID-like rather than a Fraunhofer-like interference pattern. For equivalent arms a fourth-order critical-current magnitude has the form
\begin{align}
    \mathcal{I}_{c}^{(4)}(\Phi)
    \sim
    \left|
    \mathcal{P}\mathcal{I}_c^{\mathrm{nl}}
    +2\mathcal{I}_c^{\mathrm{loc}}\cos\left(\pi\Phi/\Phi_0\right)
    \right| .
\end{align}
The local contribution alone gives the usual $\Phi_0$-periodic critical-current magnitude after taking the absolute value. Its interference with the flux-independent non-local amplitude produces alternating lobe heights and, generically, an $h/e=2\Phi_0$ fundamental period. In the exact equilibrium response, this interference pattern is further modulated by flux-induced changes of the many-body ground state. Furthermore, the critical currents $\mathcal{I}_c^{\pm}$ exhibit a major difference in the regions between the lobes, where local and non-local supercurrents interfere destructively. Precisely at these regions, the rectification efficiency is higher. A line-cut along Fig.~\ref{fig:3} for $\epsilon_1=-7.7\Delta_b$ and $\epsilon_1=-5\Delta_b$ are shown in Fig.~\ref{fig:7}, where we can see explicitly the different many body branches that contribute to $\mathcal{I}_c^\pm$ and that give rise to large variation of diode efficiencies.

Inside the $(1,1)$ sector, triplet-dominated lobes $(S=1)$ are generally reduced relative to singlet-dominated lobes $(S=0)$. Along the representative cut $\epsilon_1=-5\Delta_b$, the resulting rectification is only a few percent, although the broader spin-crossover regions in the flux--gate maps reach efficiencies of order $10$--$15\%$. A much stronger response appears close to the charge-sector boundaries, as illustrated by the line-cut $\epsilon_1=-7.7\Delta_b\simeq-U$, where the two sets of critical current lobes have different widths resulting in $|\eta|$ approaching $40\%$. The microscopic distinction between these two regimes is resolved below using the gate--interaction map and the corresponding phase-dependent spectra, see Sec.~\ref{app:additional-flux-cuts} for further details.

\subsection{Many-body origin of the two rectification regimes}

\begin{figure}[t]
    \centering
    \includegraphics[width=\linewidth,height=0.25\textwidth]{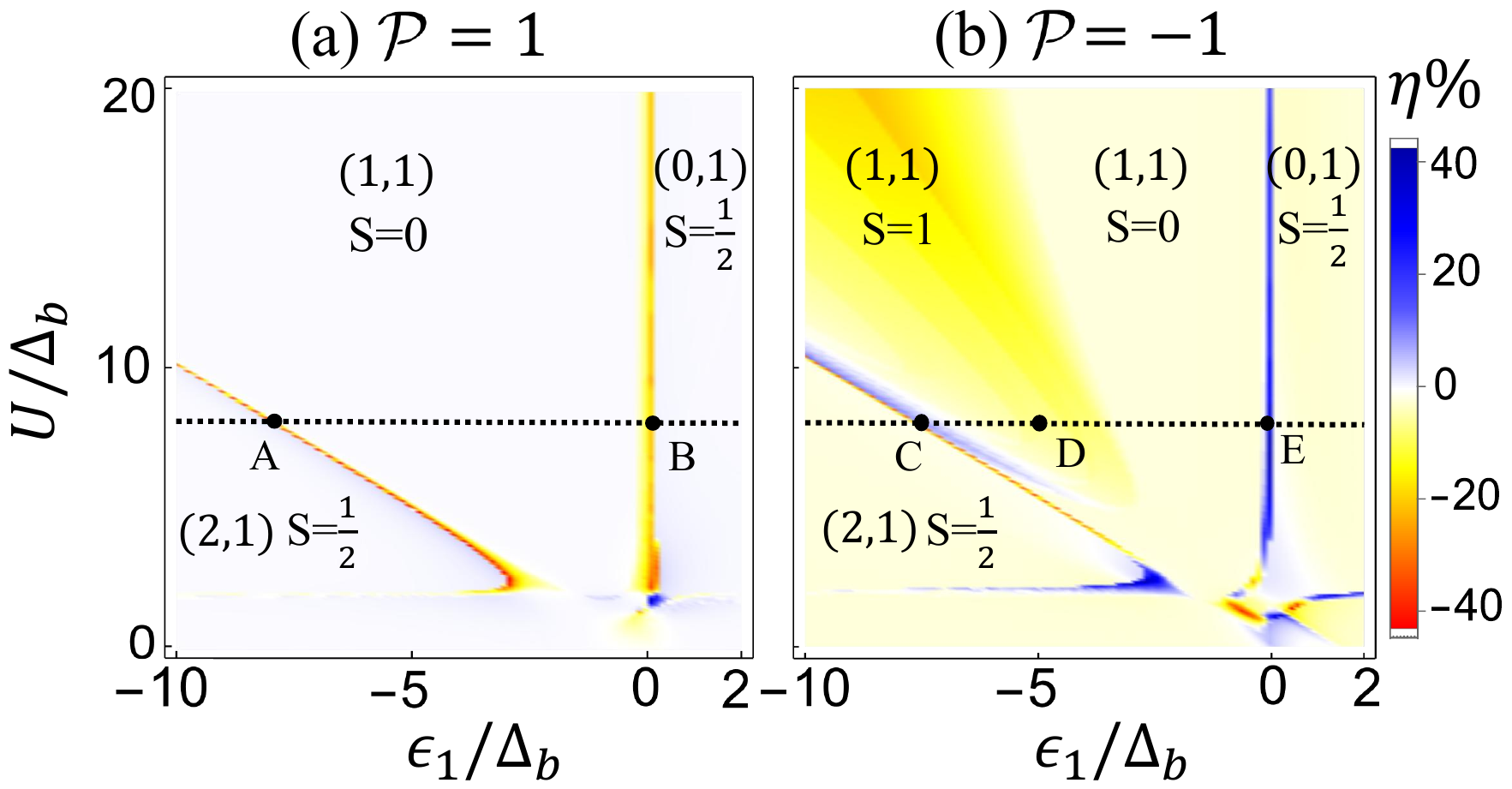}
    \caption{Josephson-diode efficiency in the $(\epsilon_1,U)$ plane for (a) even and (b) odd tunnel parity at fixed $\Phi/\Phi_0=0.76$. Labels indicate the dominant charge and total-spin character of the ground-state segments associated with the critical-current extrema. The dashed horizontal line at $U=8\Delta_b$ and points A--E specify the cuts analyzed in Fig.~\ref{fig:5}. We use $\epsilon_2=-1.5\Delta_b$ and $t_{L1}=|t_{L2}|=t_{R1}=t_{R2}=0.5\Delta_b$. 
    }
    \label{fig:4}
\end{figure}

The phase diagrams as a function of $U$ and $\epsilon_{1}$ at fixed $\Phi/\Phi_0=0.76$ are shown in Fig.~\ref{fig:4}. In the atomic limit, dot 1 is predominantly empty for $\epsilon_1>0$, singly occupied for $-U<\epsilon_1<0$, and doubly occupied for $\epsilon_1<-U$. The central $(1,1)$ sector is therefore bounded approximately by $\epsilon_1=0$ and $\epsilon_1=-U$. Finite tunneling shifts and broadens these boundaries and admixes neighboring charge configurations. Since $\epsilon_2=-1.5\Delta_b$ is fixed, the assignment $N_2=1$ applies in the atomic limit only for $U>1.5\Delta_b$.

\begin{figure*}[t]
    \centering
    \includegraphics[width=\linewidth,height=0.4\textwidth]{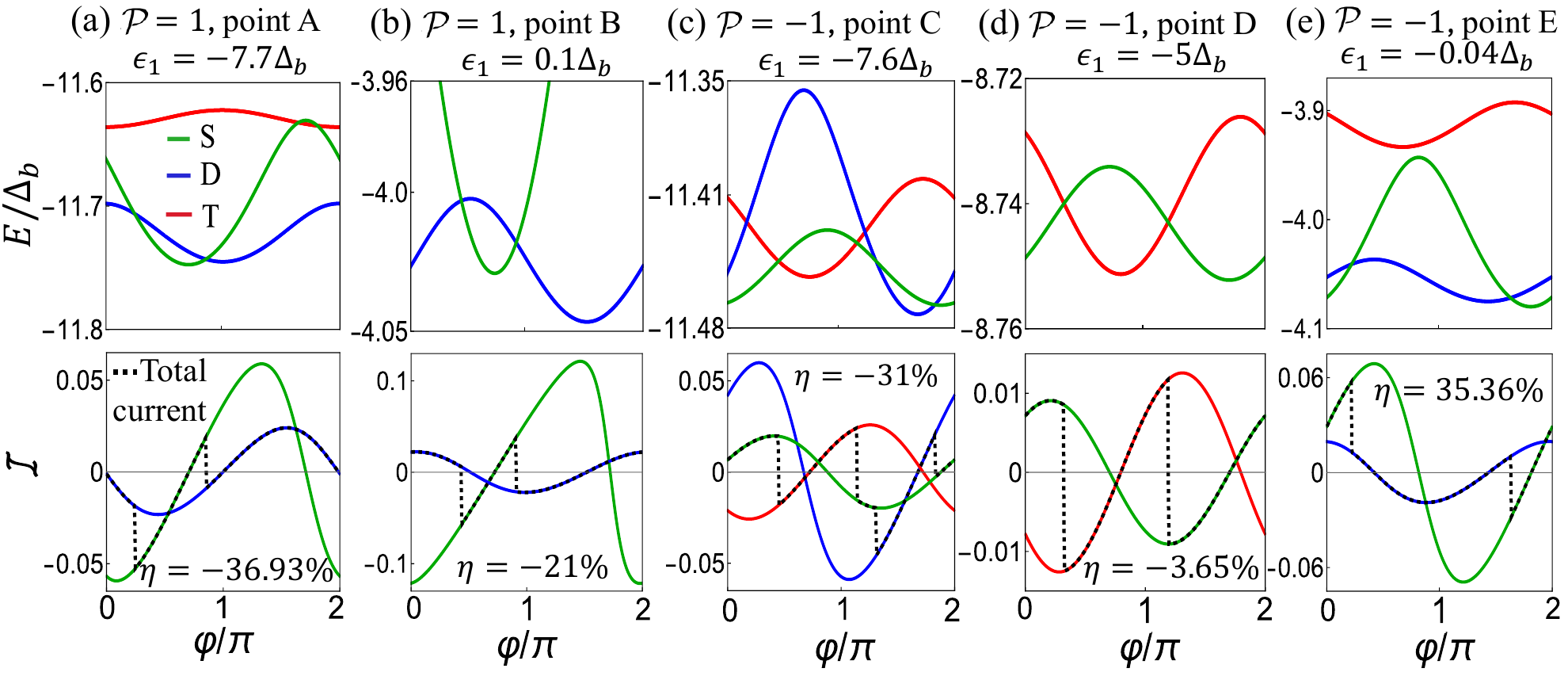}
    \caption{Lowest phase-dependent many-body energies (upper row) as well as branch resolved and total Josephson current (lower row) at the five points marked in Fig.~\ref{fig:4}. Green, blue, and red curves denote branches with singlet, doublet, and triplet character, respectively.
    The black dashed curve presents the total Josephson current and show the explicit many body branch contributing to the critical currents.
    All panels use $U=8\Delta_b$ and $\Phi/\Phi_0=0.76$; the values of $\epsilon_1$ are indicated above the panels. Other parameters are the same as in Fig.~\ref{fig:4}.
    }
    \label{fig:5}
\end{figure*}

The first rectification regime follows the narrow ridges near the $(2,1)\leftrightarrow(1,1)$ and $(1,1)\leftrightarrow(0,1)$ charge-sector boundaries. The doublet branches on the $(2,1)$ and $(0,1)$ sides are dominated by local virtual processes, whereas the competing singlet or triplet branches in the $(1,1)$ sector contain substantial non-local pair-transfer and exchange contributions. Their phase and flux dispersions are therefore markedly different. As $\varphi$ is varied, a crossing can lie close to one critical phase difference but far from the other, truncating one current extremum while leaving the opposite extremum comparatively intact. This mechanism is illustrated by points A--C and E in Fig.~\ref{fig:5}: depending on the position of the crossing, the representative efficiencies range from about $20\%$ to nearly $40\%$. A related singlet--doublet branch-selection mechanism was discussed in Ref.~\onlinecite{Zhang2026}. Here, orbital flux and tunnel parity additionally determine whether the competing $(1,1)$ branch is predominantly singlet or triplet.

The second rectification regime is associated with singlet--triplet competition within the $(1,1)$ sector and appears only in the strongly interacting regime. For the representative orbital flux $\Phi/\Phi_0=0.76$,
this regime is most clearly visible for odd tunnel parity, $\mathcal{P}=-1$, as shown in Fig.~\ref{fig:4}(b). It is not restricted to this parity, however: for $\mathcal{P}=1$, an analogous singlet--triplet transition appears in the complementary flux intervals shown in Fig.~\ref{fig:3}. Interestingly, in this regime the finite rectification region broadens and its value increases with increasing $U$, reaching values of order $10$--$15\%$ in Fig.~\ref{fig:4}(b). This is consistent with its origin in interaction-induced singlet--triplet exchange.

Fig.~\ref{fig:5}(d) reveals the key distinction from the charge-crossover regime. Both critical-current extrema are reached on triplet-like portions of the equilibrium ground-state envelope. Nevertheless, the neighboring singlet crossings delimit the phase interval over which the triplet remains the ground state and reshape the positive and negative triplet-current extrema unequally. The rectification must therefore be attributed to the complete singlet--triplet ground-state envelope rather than to an isolated triplet branch. Variations of the tunneling amplitudes modify this mechanism quantitatively rather than introducing a new one. As shown in  Figs.~\ref{fig:9}--\ref{fig:12}, changing the overall coupling or the left--right asymmetry shifts and broadens the charge- and spin-sector crossings, producing a generally nonmonotonic dependence of $\eta$ on the tunnel parameters. Further details are provided in Apps.~\ref{app:tunnel} and \ref{app:tunnel_asy}.

Although the preceding analysis focuses on orbital-flux-induced rectification, comparable diode efficiencies can also be obtained at zero orbital flux through a microscopically distinct mechanism involving a Zeeman shift of the quantum dot spin states and an asymmetrically distributed spin-orbit effect on the tunneling amplitudes. The corresponding gate-dependent maps, many-body spectra, and coupling dependence are presented in App.~\ref{app:Zeeman_Rashba}.

\subsection{Thermal suppression and enhancement of rectification}

Finite temperature replaces the zero-temperature selection of the lowest many-body branch by the statistical average 
\begin{align}
   I(\varphi)
    =\frac{2e}{\hbar}\sum_n p_n
    \langle n| \frac{\partial H}{\partial\varphi}|n\rangle.
\end{align}
Thus, the resulting thermal correction depends on both the phase-dependent energy separations between competing states and the supercurrents they carry.

\begin{figure}
    \centering
    \includegraphics[width=\linewidth,height=0.42\textwidth]{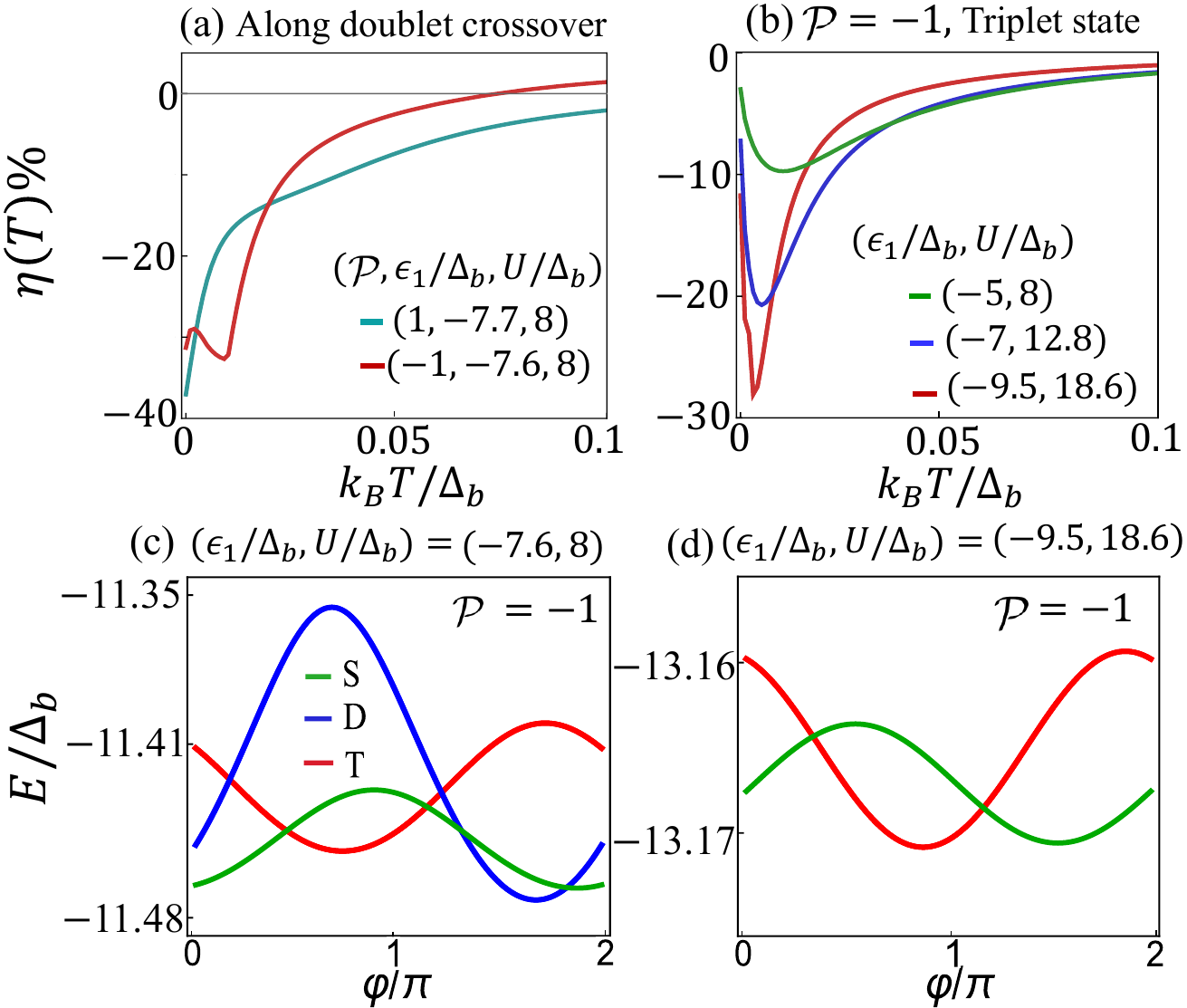}
    \caption{Temperature dependence of the Josephson-diode response. (a) Diode efficiency $\eta(T)$ at the charge-sector crossovers marked by point A for even tunnel parity $(\mathcal P,\epsilon_1/\Delta_b,U/\Delta_b)=(1,-7.7,8)$ and point C for odd tunnel parity $(-1,-7.6,8)$ in Fig.~\ref{fig:4}. (b) Diode efficiency within the triplet-dominated $(1,1)$ sector at $\mathcal P=-1$ for the three indicated pairs $(\epsilon_1/\Delta_b,U/\Delta_b)$. The parameter set $(-5,8)$ corresponds to point D in Fig.~\ref{fig:4}(b). Common parameters are $\epsilon_2=-1.5\Delta_b$, $\Phi/\Phi_0=0.76$, and $t_{L1}=|t_{L2}|=t_{R1}=t_{R2}=0.5\Delta_b$.
    (c)-(d): Lowest phase-dependent many-body energies at $\mathcal{P}=-1$. Green, blue, and red curves denote singlet, doublet, and triplet branches, respectively.}
    \label{fig:6}
\end{figure}

In the limit where temperatures exceed the bandwidth of the relevant low-energy manifold, thermal averaging reduces the phase-dependent branch selection responsible for rectification, leading to the suppression observed in Fig.~\ref{fig:6}. At intermediate temperatures, however, $|\eta|$ can either decrease monotonically or initially increase before decreasing, as illustrated in Fig.~\ref{fig:6}(a,b). For the parameters shown, temperature enhances $|\eta|$ in the triplet-dominated sector, with the efficiency approaching $30\%$, whereas it suppresses $|\eta|$ in the doublet--singlet and doublet--triplet regimes.

To understand these different responses, we examine the temperature derivative of the supercurrent. For temperature-independent Hamiltonian parameters
\begin{align}
\frac{\partial I(\varphi,T)}{\partial T}
=
\frac{1}{k_{\mathrm B}T^2}
\operatorname{Cov}_{\varphi,T}(E,I),
\label{eq:current-covariance}
\end{align}
where the energy-current covariance is
\begin{align}
\operatorname{Cov}_{\varphi,T}(E,I)
=
\sum_n p_n
\left(E_n-\langle E \rangle_T\right)
\left(I_n-\langle I\rangle_T\right),
\label{eq:cov-def}
\end{align}
with $\langle E\rangle_T=\sum_n p_nE_n$ and $\langle I\rangle_T=\sum_n p_nI_n$. %All quantities are evaluated at the same phase and temperature.

The covariance expresses how thermal redistribution connects differences in energy with differences in the current carried by the states. Increasing temperature transfers occupation toward higher-energy states, and the current changes according to whether these states carry larger or smaller sign-dependent currents than those losing occupation. As the pairwise form in Eq.~\eqref{eq:cov-def} shows, energy and current differences with the same sign contribute positively to $\partial_T I$, while opposite signs contribute negatively. Their importance is controlled jointly by the energy separation, the current contrast, and products of thermal weights $p_np_m$. In this way, large energy separations contribute little when the higher state is thermally inaccessible. Similarly, thermal redistribution between states carrying nearly equal currents has little effect on the supercurrent.

The unequal action of this redistribution at the positive and negative critical phase differences determines whether the rectification magnitude increases or decreases. At temperatures where the global critical currents vary smoothly, the condition for $d|\eta|/dT>0$ imposes
\begin{align}
\Delta I_{\mathrm c}
\left(
I_{\mathrm c}^-\mathcal C_+
+
I_{\mathrm c}^+\mathcal C_-
\right)>0,
\label{eq:thermalcond}
\end{align}
where $\mathcal C_\pm=\operatorname{Cov}_{\varphi_\pm(T),T}(E,I)$, $\Delta I_{\mathrm c}=I_{\mathrm c}^+-I_{\mathrm c}^-$ and $\varphi_{\pm}$ denotes the critical phase differences for $I_{c}^{\pm}$. This condition applies to either diode polarity and requires the favored critical current to increase relative to the unfavored one.

The enhancement observed in the triplet-dominated regime is therefore consistent with thermal occupation of nearby singlet states modifying the two critical currents unequally. More generally, temperature can either strengthen or reduce their relative imbalance, producing a nonmonotonic diode response. The direction of this change is controlled by the energy--current correlations at the global critical phase differences and cannot be inferred solely from the spin character of the zero-temperature ground state.

%------------------------------------

%------------------------------------
%----------------------------------
\section{Conclusions}\label{sec:conclusions}
%----------------------------------

We have investigated equilibrium Josephson nonreciprocity in a strongly interacting parallel double-quantum-dot junction using exact diagonalization within the zero-bandwidth approximation, complemented by a fourth-order decomposition of the virtual tunneling processes. The central organizing principle is the phase-dependent reconstruction of the many-body ground-state envelope. Broken time-reversal and arm-exchange symmetries permit a diode response, but do not generate it by themselves: \emph{at fourth order}, an adiabatically followed individual branch contains only a shifted first harmonic and is therefore reciprocal. Nonreciprocity appears when singlet, triplet, and doublet branches with inequivalent phase dispersions compete for the ground state, while higher harmonics retained by the exact diagonalization can further skew the current--phase relation. 

The gauge-invariant formulation separates the roles of the different transport processes. At fixed mean superconducting phase, the non-local Josephson contribution is independent of the orbital flux, the two local channels acquire opposite flux-induced phase shifts, and non-local cotunneling enters the singlet--triplet exchange through $\mathcal{P}\cos(\pi\Phi/\Phi_0)$. The tunnel parity therefore selects complementary singlet- and triplet-dominated flux intervals, while interference between flux-independent non-local transfer and the shifted local amplitudes produces alternating critical-current lobes with a generic $h/e=2\Phi_0$ period. Within this structure, we find two rectification regimes. Near the $(2,1)\leftrightarrow(1,1)$ and $(1,1)\leftrightarrow(0,1)$ boundaries, competition between doublet and singlet or triplet branches yields efficiencies approaching $30$--$40\%$. A broader response of order $10$--$15\%$ occurs inside the triplet-dominated $(1,1)$ sector. There, both critical-current extrema may lie on triplet-like segments, yet nearby singlet crossings delimit the interval over which the triplet is the ground state and modify the two extrema unequally. Thus, different branches do not need to carry the two critical currents. In turn, it is sufficient that nearby competing states asymmetrically reshape the equilibrium envelope. 

Temperature provides an additional control of the diode response. For the parameters studied, thermal averaging suppresses nonreciprocity near the charge-sector boundaries, whereas it can enhance $|\eta|$ in the triplet-dominated $(1,1)$ sector, with efficiencies approaching $30\%$. This behavior originates from the phase-dependent redistribution of occupation among many-body states carrying different supercurrents. The resulting thermal correction is governed by the energy--current covariance, which combines energy separations, current contrasts, and thermal weights. Its unequal action at the two critical phase differences can strengthen or reduce the relative imbalance between the critical currents. At higher temperatures, broader thermal occupation weakens the branch-selection mechanism and suppresses the rectification observed here. The thermal response therefore reflects the competing branches and their current dispersions, neither the ground-state spin nor its multiplicity alone determines whether temperature enhances or suppresses nonreciprocity.

The singlet-triplet competition relevant to the present mechanism is supported beyond the zero-bandwidth approximation. For the same parallel-DQD Josephson geometry, Ref.~\onlinecite{Probst2016} obtained both the transition and the cotunneling stabilization of the triplet in a fourth-order wide-band treatment of continuum BCS reservoirs. This supports the microscopic basis of the competing spin sectors, while a dedicated continuum calculation is needed to benchmark the finite-flux phase boundaries, current amplitudes, and diode efficiencies. Additional interdot couplings and constraints on parity or spin relaxation may further modify the response. Our results identify gate voltage, orbital flux, tunnel parity, and temperature as complementary controls of equilibrium Josephson nonreciprocity arising from the competition and thermal occupation of interacting many-body states as well as the influence of strong Coulomb interaction.

\begin{acknowledgments}
D.\ D.\ and P.\ R.\ gratefully acknowledge funding by the Deutsche Forschungsgemeinschaft (DFG, German Research Foundation) under Germany’s Excellence Strategy –EXC-2123/2 QuantumFrontiers – 390837967. F.\ D.\ acknowledges funding support from the Deutsche Forschungsgemeinschaft (DFG, German Research Foundation) under Germany’s Excellence Strategy through the W\"urzburg-Dresden Cluster of Excellence ctd.qmat (EXC 2147, Project ID 390858490) as well as through the Collaborative Research
Center SFB 1170 ToCoTronics (Project ID 258499086).
\end{acknowledgments}

\bibliography{ref}

@article{Recher2002,
  title = {Superconductor coupled to two Luttinger liquids as an entangler for electron spins},
  author = {Recher, Patrik and Loss, Daniel},
  journal = {Phys. Rev. B},
  volume = {65},
  issue = {16},
  pages = {165327},
  numpages = {13},
  year = {2002},
  month = {Apr},
  publisher = {American Physical Society},
  doi = {10.1103/PhysRevB.65.165327},
  url = {https://link.aps.org/doi/10.1103/PhysRevB.65.165327}
}

@article{Recher2001,
  title = {Andreev tunneling, Coulomb blockade, and resonant transport of nonlocal spin-entangled electrons},
  author = {Recher, Patrik and Sukhorukov, Eugene V. and Loss, Daniel},
  journal = {Phys. Rev. B},
  volume = {63},
  issue = {16},
  pages = {165314},
  numpages = {11},
  year = {2001},
  month = {Apr},
  publisher = {American Physical Society},
  doi = {10.1103/PhysRevB.63.165314},
  url = {https://link.aps.org/doi/10.1103/PhysRevB.63.165314}
}

@article{Probst2016,
  title = {Signatures of nonlocal Cooper-pair transport and of a singlet-triplet transition in the critical current of a double-quantum-dot Josephson junction},
  author = {Probst, B. and Dom\'{\i}nguez, F. and Schroer, A. and Yeyati, A. Levy and Recher, P.},
  journal = {Phys. Rev. B},
  volume = {94},
  issue = {15},
  pages = {155445},
  numpages = {11},
  year = {2016},
  month = {Oct},
  publisher = {American Physical Society},
  doi = {10.1103/PhysRevB.94.155445},
  url = {https://link.aps.org/doi/10.1103/PhysRevB.94.155445}
}

@misc{Zhang2026,
      title={Many-body Josephson diode effect in superconducting quantum interferometers}, 
      author={Zelei Zhang and Jianxiong Zhai and Yi Zhang and Jiawei Yan},
      year={2026},
      eprint={2603.25558},
      archivePrefix={arXiv},
      url={https://arxiv.org/abs/2603.25558}
      }

@article{Affleck2000,
  title = {Andreev scattering and Josephson current in a one-dimensional electron liquid},
  author = {Affleck, Ian and Caux, Jean-S\'ebastien and Zagoskin, Alexandre M.},
  journal = {Phys. Rev. B},
  volume = {62},
  issue = {2},
  pages = {1433--1445},
  numpages = {0},
  year = {2000},
  month = {Jul},
  publisher = {American Physical Society},
  doi = {10.1103/PhysRevB.62.1433},
  url = {https://link.aps.org/doi/10.1103/PhysRevB.62.1433}
}

@article{Grove2018,
author={Grove-Rasmussen, K.
and Steffensen, G.
and Jellinggaard, A.
and Madsen, M. H.
and {\v{Z}}itko, R.
and Paaske, J.
and Nyg{\aa}rd, J.},
title={Yu--Shiba--Rusinov screening of spins in double quantum dots},
journal={Nature Communications},
year={2018},
month={Jun},
day={18},
volume={9},
number={1},
pages={2376},
issn={2041-1723},
doi={10.1038/s41467-018-04683-x},
url={https://doi.org/10.1038/s41467-018-04683-x}
}

@article{Zalom2024,
  title = {Double quantum dot Andreev molecules: Phase diagrams and critical evaluation of effective models},
  author = {Zalom, Peter and Wrze\ifmmode \acute{s}\else \'{s}\fi{}niewski, Kacper and Novotn\'y, Tom\'a\ifmmode \check{s}\else \v{s}\fi{} and Weymann, Ireneusz},
  journal = {Phys. Rev. B},
  volume = {110},
  issue = {13},
  pages = {134506},
  numpages = {19},
  year = {2024},
  month = {Oct},
  publisher = {American Physical Society},
  doi = {10.1103/PhysRevB.110.134506},
  url = {https://link.aps.org/doi/10.1103/PhysRevB.110.134506}
}

@article{Yi2024,
  title = {Spin-resolved Yu-Shiba-Rusinov states and bistable phase caused by side-coupled Majorana bound states in quantum-dot Josephson junctions},
  author = {Yi, Guang-Yu and Ma, Meng-Ya and Gong, Tong and Jiang, Cui and Gong, Wei-Jiang},
  journal = {Phys. Rev. B},
  volume = {109},
  issue = {16},
  pages = {165433},
  numpages = {20},
  year = {2024},
  month = {Apr},
  publisher = {American Physical Society},
  doi = {10.1103/PhysRevB.109.165433},
  url = {https://link.aps.org/doi/10.1103/PhysRevB.109.165433}
}

@article{Debbarma2023,
  title = {Josephson Junction $\ensuremath{\pi}\text{\ensuremath{-}}0$ Transition Induced by Orbital Hybridization in a Double Quantum Dot},
  author = {Debbarma, Rousan and Tsintzis, Athanasios and Aspegren, Markus and Souto, Rub\'en Seoane and Lehmann, Sebastian and Dick, Kimberly and Leijnse, Martin and Thelander, Claes},
  journal = {Phys. Rev. Lett.},
  volume = {131},
  issue = {25},
  pages = {256001},
  numpages = {6},
  year = {2023},
  month = {Dec},
  publisher = {American Physical Society},
  doi = {10.1103/PhysRevLett.131.256001},
  url = {https://link.aps.org/doi/10.1103/PhysRevLett.131.256001}
}

@article{Souto2022,
  title = {Josephson Diode Effect in Supercurrent Interferometers},
  author = {Souto, Rub\'en Seoane and Leijnse, Martin and Schrade, Constantin},
  journal = {Phys. Rev. Lett.},
  volume = {129},
  issue = {26},
  pages = {267702},
  numpages = {6},
  year = {2022},
  month = {Dec},
  publisher = {American Physical Society},
  doi = {10.1103/PhysRevLett.129.267702},
  url = {https://link.aps.org/doi/10.1103/PhysRevLett.129.267702}
}

@article{Roy2025,
  title = {Floquet-engineered diode performance of a topological Josephson junction composed of two Kitaev chains coupled via a quantum dot},
  author = {Roy, Koustav and Paul, Gourab and Debnath, Debika and Bhattacharyya, Kuntal and Basu, Saurabh},
  journal = {Phys. Rev. B},
  volume = {112},
  issue = {12},
  pages = {125407},
  numpages = {18},
  year = {2025},
  month = {Sep},
  publisher = {American Physical Society},
  doi = {10.1103/69jq-rcsb}
}

@article{Debnath2024,
  title = {Gate-tunable Josephson diode effect in Rashba spin-orbit coupled quantum dot junctions},
  author = {Debnath, Debika and Dutta, Paramita},
  journal = {Phys. Rev. B},
  volume = {109},
  issue = {17},
  pages = {174511},
  numpages = {11},
  year = {2024},
  month = {May},
  publisher = {American Physical Society},
  doi = {10.1103/PhysRevB.109.174511},
  url = {https://link.aps.org/doi/10.1103/PhysRevB.109.174511}
}

@article{Debnath2025,
  author  = {Debnath, Debika and Dutta, Paramita},
  title   = {Field-free Josephson diode effect in interacting chiral quantum dot junctions},
  journal = {Journal of Physics: Condensed Matter},
  volume  = {37},
  number  = {17},
  pages   = {175301},
  year    = {2025},
  doi     = {10.1088/1361-648X/adbeaf}
}

@article{Nikodem2025,
  author  = {Nikodem, Ella and Schluck, Jakob and Geier, Max and Papaj, Micha{\l} and Legg, Henry F. and Feng, Junya and Bagchi, Mahasweta and Fu, Liang and Ando, Yoichi},
  title   = {Tunable superconducting diode effect in a topological nano-{SQUID}},
  journal = {Science Advances},
  volume  = {11},
  number  = {38},
  pages   = {eadw4898},
  year    = {2025},
  doi     = {10.1126/sciadv.adw4898}
}

@article{Greco2024,
  author  = {Greco, Angelo and Pichard, Quentin and Strambini, Elia and Giazotto, Francesco},
  title   = {Double loop dc-{SQUID} as a tunable {Josephson} diode},
  journal = {Applied Physics Letters},
  volume  = {125},
  number  = {7},
  pages   = {072601},
  year    = {2024},
  doi     = {10.1063/5.0211021},
  url     = {https://doi.org/10.1063/5.0211021}
}

@article{Wang2011,
  title = {Interference and Switching of Josephson Current Carried by Nonlocal Spin-Entangled Electrons in a SQUID-Like System with Quantum Dots},
  author = {Wang, Zhi and Hu, Xiao},
  journal = {Phys. Rev. Lett.},
  volume = {106},
  issue = {3},
  pages = {037002},
  numpages = {4},
  year = {2011},
  month = {Jan},
  publisher = {American Physical Society},
  doi = {10.1103/PhysRevLett.106.037002},
  url = {https://link.aps.org/doi/10.1103/PhysRevLett.106.037002}
}

@article{Dominguez2024a,
  title = {Fraunhofer pattern in the presence of Majorana zero modes},
  author = {Dominguez, F. and Novik, E. G. and Recher, P.},
  journal = {Phys. Rev. Res.},
  volume = {6},
  issue = {2},
  pages = {023304},
  numpages = {17},
  year = {2024},
  month = {Jun},
  publisher = {American Physical Society},
  doi = {10.1103/PhysRevResearch.6.023304},
  url = {https://link.aps.org/doi/10.1103/PhysRevResearch.6.023304}
}

@article{Haidekker2020a,
  title = {Superconducting Quantum Interference in Edge State Josephson Junctions},
  author = {Haidekker Galambos, Tam\'as and Hoffman, Silas and Recher, Patrik and Klinovaja, Jelena and Loss, Daniel},
  journal = {Phys. Rev. Lett.},
  volume = {125},
  issue = {15},
  pages = {157701},
  numpages = {6},
  year = {2020},
  month = {Oct},
  publisher = {American Physical Society},
  doi = {10.1103/PhysRevLett.125.157701},
  url = {https://link.aps.org/doi/10.1103/PhysRevLett.125.157701}
}

@article{Takeuchi2025,
  author  = {Takeuchi, Go and Eto, Mikio},
  title   = {Superconducting Diode Effect in Double Quantum Dot Device},
  journal = {Journal of the Physical Society of Japan},
  volume  = {94},
  pages   = {054701},
  year    = {2025},
  doi     = {10.7566/JPSJ.94.054701}
}

@article{Choi2000,
  title = {Spin-dependent Josephson current through double quantum dots and measurement of entangled electron states},
  author = {Choi, Mahn-Soo and Bruder, C. and Loss, Daniel},
  journal = {Phys. Rev. B},
  volume = {62},
  issue = {20},
  pages = {13569--13572},
  numpages = {0},
  year = {2000},
  month = {Nov},
  publisher = {American Physical Society},
  doi = {10.1103/PhysRevB.62.13569},
  url = {https://link.aps.org/doi/10.1103/PhysRevB.62.13569}
}

@article{Thieme2026,
	title = {Magnetic field-bias current interplay in HgTe-based three-terminal Josephson junctions},
	pages = {012},
	author = {Thieme, Josua and Himmler, Wolfgang and Dominguez, Fernando and Platero, Gloria and Hüttner, Niklas and Hartl, Stefan Felix and Richter, Elisabeth and Kozlov, Dmitriy and Mikhailov, Nikolai and Dvoretsky, Sergey Alekseevich and Weiss, Dieter},
	journal = {SciPost Phys.},
	volume = {21},
	year = {2026},
	publisher = {SciPost},
	doi = {10.21468/SciPostPhys.21.1.012},
	url = {https://scipost.org/10.21468/SciPostPhys.21.1.012}
}

@article{Vecino2003a,
  title = {Josephson current through a correlated quantum level: Andreev states and $\ensuremath{\pi}$ junction behavior},
  author = {Vecino, E. and Mart\'{\i}n-Rodero, A. and Yeyati, A. Levy},
  journal = {Phys. Rev. B},
  volume = {68},
  issue = {3},
  pages = {035105},
  numpages = {9},
  year = {2003},
  month = {Jul},
  publisher = {American Physical Society},
  doi = {10.1103/PhysRevB.68.035105},
  url = {https://link.aps.org/doi/10.1103/PhysRevB.68.035105}
}

@article{Bergeret2007a,
  title = {Josephson effect through a quantum dot array},
  author = {Bergeret, F. S. and Yeyati, A. Levy and Mart\'{\i}n-Rodero, A.},
  journal = {Phys. Rev. B},
  volume = {76},
  issue = {17},
  pages = {174510},
  numpages = {6},
  year = {2007},
  month = {Nov},
  publisher = {American Physical Society},
  doi = {10.1103/PhysRevB.76.174510},
  url = {https://link.aps.org/doi/10.1103/PhysRevB.76.174510}
}

@article{Deacon2015,
  author  = {Deacon, R. S. and Oiwa, A. and Sailer, J. and Baba, S. and Kanai, Y. and Shibata, K. and Hirakawa, K. and Tarucha, S.},
  title   = {Cooper pair splitting in parallel quantum dot Josephson junctions},
  journal = {Nature Communications},
  volume  = {6},
  pages   = {7446},
  year    = {2015},
  doi     = {10.1038/ncomms8446}
}

@article{Ando2020,
  author  = {Fuyuki Ando and Yuta Miyasaka and Tian Li and Jun Ishizuka and Tomonori Arakawa and Yoichi Shiota and Takahiro Moriyama and Youichi Yanase and Teruo Ono},
  title   = {Observation of superconducting diode effect},
  journal = {Nature},
  volume  = {584},
  pages   = {373--376},
  year    = {2020},
  doi     = {10.1038/s41586-020-2590-4}
}

@article{Baumgartner2022,
  author  = {Christian Baumgartner and Lorenz Fuchs and Andreas Costa and Simon Reinhardt and Sergei Gronin and Geoffrey C. Gardner and Tyler Lindemann and Michael J. Manfra and Paulo E. Faria Junior and Denis Kochan and Jaroslav Fabian and Nicola Paradiso and Christoph Strunk},
  title   = {Supercurrent rectification and magnetochiral effects in symmetric Josephson junctions},
  journal = {Nature Nanotechnology},
  volume  = {17},
  pages   = {39--44},
  year    = {2022},
  doi     = {10.1038/s41565-021-01009-9}
}

@article{He2023,
    author ={He, J. J. and Tanaka, Y. and Nagaosa, N.} ,
    title ={The supercurrent diode effect and nonreciprocal paraconductivity due to the chiral structure of nanotubes} ,
    journal = {Nature Communications},
    year = {2023},
    volume = {14},
    pages = {3330},
    publisher = {Nature},
    doi = {10.1038/s41467-023-39083-3},
    url = {https://doi.org/10.1038/s41467-023-39083-3}
}

@article{Legg2022,
  title = {Superconducting diode effect due to magnetochiral anisotropy in topological insulators and Rashba nanowires},
  author = {Legg, Henry F. and Loss, Daniel and Klinovaja, Jelena},
  journal = {Phys. Rev. B},
  volume = {106},
  issue = {10},
  pages = {104501},
  numpages = {8},
  year = {2022},
  month = {Sep},
  publisher = {American Physical Society},
  doi = {10.1103/PhysRevB.106.104501},
  url = {https://link.aps.org/doi/10.1103/PhysRevB.106.104501}
}

@article{Lu2023,
  title = {Tunable Josephson Diode Effect on the Surface of Topological Insulators},
  author = {Lu, Bo and Ikegaya, Satoshi and Burset, Pablo and Tanaka, Yukio and Nagaosa, Naoto},
  journal = {Phys. Rev. Lett.},
  volume = {131},
  issue = {9},
  pages = {096001},
  numpages = {7},
  year = {2023},
  month = {Aug},
  publisher = {American Physical Society},
  doi = {10.1103/PhysRevLett.131.096001},
  url = {https://link.aps.org/doi/10.1103/PhysRevLett.131.096001}
}

@article{Davydova2022,
  author  = {Margarita Davydova and Saranesh Prembabu and Liang Fu},
  title   = {Universal Josephson diode effect},
  journal = {Science Advances},
  volume  = {8},
  pages   = {eabo0309},
  year    = {2022},
  doi     = {10.1126/sciadv.abo0309}
}

@article{Pal2022,
  author  = {Banabir Pal and Anirban Chakraborty and Pranava K. Sivakumar and Margarita Davydova and Ajesh K. Gopi and Avanindra K. Pandeya and Jonas A. Krieger and Yang Zhang and Mihir Date and Sailong Ju and Noah Yuan and Niels B. M. Schr{\"o}ter and Liang Fu and Stuart S. P. Parkin},
  title   = {Josephson diode effect from Cooper pair momentum in a topological semimetal},
  journal = {Nature Physics},
  volume  = {18},
  pages   = {1228--1233},
  year    = {2022},
  doi     = {10.1038/s41567-022-01699-5}
}

@article{Fu2022,
  title={Supercurrent diode effect and finite-momentum superconductors},
  author={Yuan, Noah FQ and Fu, Liang},
  journal={Proceedings of the National Academy of Sciences},
  volume={119},
  number={15},
  pages={e2119548119},
  year={2022},
  publisher={National Acad Sciences},
  url={https://www.pnas.org/doi/10.1073/pnas.2119548119},
    doi={10.1073/pnas.2119548119}
}

@article{Yanase2022,
  title = {Intrinsic Superconducting Diode Effect},
  author = {Daido, Akito and Ikeda, Yuhei and Yanase, Youichi},
  journal = {Phys. Rev. Lett.},
  volume = {128},
  issue = {3},
  pages = {037001},
  numpages = {6},
  year = {2022},
  month = {Jan},
  publisher = {American Physical Society},
  doi = {10.1103/PhysRevLett.128.037001},
  url = {https://link.aps.org/doi/10.1103/PhysRevLett.128.037001}
}

@article{Chakraborty2025,
  title = {Perfect Superconducting Diode Effect in Altermagnets},
  author = {Chakraborty, Debmalya and Black-Schaffer, Annica M.},
  journal = {Phys. Rev. Lett.},
  volume = {135},
  issue = {2},
  pages = {026001},
  numpages = {9},
  year = {2025},
  month = {Jul},
  publisher = {American Physical Society},
  doi = {10.1103/cv8s-tk4c},
  url = {https://link.aps.org/doi/10.1103/cv8s-tk4c}
}

@article{Mondal2025,
  title = {Josephson diode effect with Andreev and Majorana bound states},
  author = {Mondal, Sayan and Fu, Pei-Hao and Cayao, Jorge},
  journal = {Phys. Rev. B},
  volume = {112},
  issue = {14},
  pages = {144506},
  numpages = {16},
  year = {2025},
  month = {Oct},
  publisher = {American Physical Society},
  doi = {10.1103/79tj-c3y4},
  url = {https://link.aps.org/doi/10.1103/79tj-c3y4}
}

@article{Pillet2023,
  title = {Josephson diode effect in Andreev molecules},
  author = {Pillet, J.-D. and Annabi, S. and Peugeot, A. and Riechert, H. and Arrighi, E. and Griesmar, J. and Bretheau, L.},
  journal = {Phys. Rev. Res.},
  volume = {5},
  issue = {3},
  pages = {033199},
  numpages = {6},
  year = {2023},
  month = {Sep},
  publisher = {American Physical Society},
  doi = {10.1103/PhysRevResearch.5.033199},
  url = {https://link.aps.org/doi/10.1103/PhysRevResearch.5.033199}
}

@article{Nadeem2023,
  author  = {Nadeem, Muhammad and Fuhrer, Michael S. and Wang, Xiaolin},
  title   = {The superconducting diode effect},
  journal = {Nature Reviews Physics},
  volume  = {5},
  pages   = {558--577},
  year    = {2023},
  doi     = {10.1038/s42254-023-00632-w}
}

@article{Tatara2026,
  title = {Josephson phase shift and diode effect due to the inverse spin Hall effect},
  author = {Tatara, Gen and Takane, Yositake and Manchon, Aurelien},
  journal = {Phys. Rev. B},
  pages = {},
  year = {2026},
  month = {Aug},
  publisher = {American Physical Society},
  doi = {10.1103/pyhy-lc7r},
  url = {https://link.aps.org/doi/10.1103/pyhy-lc7r}
}

@article{Chen2024,
  author  = {Chen, Shaowen and Park, Seunghyun and Vool, Uri and Maksimovic, Nikola and Broadway, David A. and Flaks, Mykhailo and Zhou, Tony X. and Maletinsky, Patrick and Stern, Ady and Halperin, Bertrand I. and Yacoby, Amir},
  title   = {Current induced hidden states in Josephson junctions},
  journal = {Nature Communications},
  volume  = {15},
  pages   = {8059},
  year    = {2024},
  doi     = {10.1038/s41467-024-52271-z}
}

@Article{Wu2022,
author={Wu, Heng
and Wang, Yaojia
and Xu, Yuanfeng
and Sivakumar, Pranava K.
and Pasco, Chris
and Filippozzi, Ulderico
and Parkin, Stuart S. P.
and Zeng, Yu-Jia
and McQueen, Tyrel
and Ali, Mazhar N.},
title={The field-free Josephson diode in a van der Waals heterostructure},
journal={Nature},
year={2022},
month={Apr},
day={01},
volume={604},
number={7907},
pages={653-656},
issn={1476-4687},
doi={10.1038/s41586-022-04504-8},
url={https://doi.org/10.1038/s41586-022-04504-8}
}

@article{Costa2023,
  title = {Microscopic study of the Josephson supercurrent diode effect in Josephson junctions based on two-dimensional electron gas},
  author = {Costa, Andreas and Fabian, Jaroslav and Kochan, Denis},
  journal = {Phys. Rev. B},
  volume = {108},
  issue = {5},
  pages = {054522},
  numpages = {13},
  year = {2023},
  month = {Aug},
  publisher = {American Physical Society},
  doi = {10.1103/PhysRevB.108.054522},
  url = {https://link.aps.org/doi/10.1103/PhysRevB.108.054522}
}

@article{Narita2022,
  title={Field-free superconducting diode effect in noncentrosymmetric superconductor/ferromagnet multilayers},
  author={Narita, Hideki and Ishizuka, Jun and Kawarazaki, Ryo and Kan, Daisuke and Shiota, Yoichi and Moriyama, Takahiro and Shimakawa, Yuichi and Ognev, Alexey V and Samardak, Alexander S and Yanase, Youichi and others},
  journal={Nature Nanotechnology},
  volume={17},
  number={8},
  pages={823--828},
  year={2022},
  publisher={Nature Publishing Group UK London},
  url={https://www.nature.com/articles/s41565-022-01159-4}
}

@article{Yerin2025,
  author  = {Yerin, Yuriy and Drechsler, Stefan-Ludwig and Varlamov, A. A. and Giazotto, Francesco and Cuoco, Mario},
  title   = {Supercurrent diode effect in Josephson interferometers with multiband superconductors},
  journal = {Communications Physics},
  volume  = {8},
  pages   = {356},
  year    = {2025},
  doi     = {10.1038/s42005-025-02253-4}
}

@article{Yerin2024,
  title = {Supercurrent rectification with time-reversal symmetry broken multiband superconductors},
  author = {Yerin, Yuriy and Drechsler, Stefan-Ludwig and Varlamov, A. A. and Cuoco, Mario and Giazotto, Francesco},
  journal = {Phys. Rev. B},
  volume = {110},
  issue = {5},
  pages = {054501},
  numpages = {12},
  year = {2024},
  month = {Aug},
  publisher = {American Physical Society},
  doi = {10.1103/PhysRevB.110.054501},
  url = {https://link.aps.org/doi/10.1103/PhysRevB.110.054501}
}

@Article{Bauriedl2022,
author={Bauriedl, Lorenz
and B{\"a}uml, Christian
and Fuchs, Lorenz
and Baumgartner, Christian
and Paulik, Nicolas
and Bauer, Jonas M.
and Lin, Kai-Qiang
and Lupton, John M.
and Taniguchi, Takashi
and Watanabe, Kenji
and Strunk, Christoph
and Paradiso, Nicola},
title={Supercurrent diode effect and magnetochiral anisotropy in few-layer NbSe2},
journal={Nature Communications},
year={2022},
month={Jul},
day={23},
volume={13},
number={1},
pages={4266},
issn={2041-1723},
doi={10.1038/s41467-022-31954-5},
url={https://doi.org/10.1038/s41467-022-31954-5}
}

@article{Ciaccia2023,
  author  = {Carlo Ciaccia and Roy Haller and Asbj{\o}rn C. C. Drachmann and Tyler Lindemann and Michael J. Manfra and Constantin Schrade and Christian Sch{\"o}nenberger},
  title   = {Gate-tunable Josephson diode in proximitized {InAs} supercurrent interferometers},
  journal = {Physical Review Research},
  volume  = {5},
  pages   = {033131},
  year    = {2023},
  doi     = {10.1103/PhysRevResearch.5.033131}
}

@article{Matsuo2023,
  author  = {Sadashige Matsuo and Takaya Imoto and Tomohiro Yokoyama and Yosuke Sato and Tyler Lindemann and Sergei Gronin and Geoffrey C. Gardner and Michael J. Manfra and Seigo Tarucha},
  title   = {Josephson diode effect derived from short-range coherent coupling},
  journal = {Nature Physics},
  volume  = {19},
  pages   = {1636--1641},
  year    = {2023},
  doi     = {10.1038/s41567-023-02144-x}
}

@article{Pankratova2020,
  author  = {Natalia Pankratova and Hanho Lee and Roman Kuzmin and Kaushini Wickramasinghe and William Mayer and Joseph Yuan and Maxim G. Vavilov and Javad Shabani and Vladimir E. Manucharyan},
  title   = {Multiterminal Josephson Effect},
  journal = {Physical Review X},
  volume  = {10},
  pages   = {031051},
  year    = {2020},
  doi     = {10.1103/PhysRevX.10.031051}
}

@article{HuamaniCorrea2024,
  author  = {Jorge Huamani Correa and Micha{\l} P. Nowak},
  title   = {Theory of universal diode effect in three-terminal Josephson junctions},
  journal = {SciPost Physics},
  volume  = {17},
  pages   = {037},
  year    = {2024},
  doi     = {10.21468/SciPostPhys.17.2.037}
}

@article{Yang2025,
author={Yang, Hung-Yu
and Cuozzo, Joseph J.
and Bokka, Anand Johnson
and Qiu, Gang
and Eckberg, Christopher
and Lyu, Yanfeng
and Huyan, Shuyuan
and Chu, Ching-Wu
and Watanabe, Kenji
and Taniguchi, Takashi
and Wang, Kang L.},
title={Field-resilient supercurrent diode in a multiferroic Josephson junction},
journal={Nature Communications},
year={2025},
month={Oct},
day={21},
volume={16},
number={1},
pages={9287},
issn={2041-1723},
doi={10.1038/s41467-025-63698-3},
url={https://doi.org/10.1038/s41467-025-63698-3}
}

@article{Kokkeler2022,
  title = {Field-free anomalous junction and superconducting diode effect in spin-split superconductor/topological insulator junctions},
  author = {Kokkeler, T. H. and Golubov, A. A. and Bergeret, F. S.},
  journal = {Phys. Rev. B},
  volume = {106},
  issue = {21},
  pages = {214504},
  numpages = {8},
  year = {2022},
  month = {Dec},
  publisher = {American Physical Society},
  doi = {10.1103/PhysRevB.106.214504},
  url = {https://link.aps.org/doi/10.1103/PhysRevB.106.214504}
}

@article{Maiani2023,
  title = {Nonsinusoidal current-phase relations in semiconductor--superconductor-- ferromagnetic insulator devices},
  author = {Maiani, Andrea and Flensberg, Karsten and Leijnse, Martin and Schrade, Constantin and Vaitiek\ifmmode \dot{e}\else \.{e}\fi{}nas, Saulius and Seoane Souto, Rub\'en},
  journal = {Phys. Rev. B},
  volume = {107},
  issue = {24},
  pages = {245415},
  numpages = {12},
  year = {2023},
  month = {Jun},
  publisher = {American Physical Society},
  doi = {10.1103/PhysRevB.107.245415},
  url = {https://link.aps.org/doi/10.1103/PhysRevB.107.245415}
}

@article{Ghosh2024,
author={Ghosh, Sanat
and Patil, Vilas
and Basu, Amit
and {Kuldeep}
and Dutta, Achintya
and Jangade, Digambar A.
and Kulkarni, Ruta
and Thamizhavel, A.
and Steiner, Jacob F.
and von Oppen, Felix
and Deshmukh, Mandar M.},
title={High-temperature Josephson diode},
journal={Nature Materials},
year={2024},
month={May},
day={01},
volume={23},
number={5},
pages={612-618},
issn={1476-4660},
doi={10.1038/s41563-024-01804-4},
url={https://doi.org/10.1038/s41563-024-01804-4}
}

@article{Debnath2026,
  title = {Spin polarization and diode effect in thermoelectric current through altermagnet-based superconductor heterostructures},
  author = {Debnath, Debika and Saha, Arijit and Dutta, Paramita},
  journal = {Phys. Rev. B},
  volume = {113},
  issue = {10},
  pages = {104508},
  numpages = {17},
  year = {2026},
  month = {Mar},
  publisher = {American Physical Society},
  doi = {10.1103/jd41-26xr},
  url = {https://link.aps.org/doi/10.1103/jd41-26xr}
}

\FloatBarrier
\appendix

%------------------------------------------------------
\section{Gauge-invariant phase structure and fourth-order effective exchange}
\label{app:fourth_order_exchange}
%------------------------------------------------------
This appendix derives the phase, flux, tunnel-parity, and spin structure of the fourth-order virtual processes used to interpret the exact-diagonalization results.  The perturbative construction is not used to replace the exact zero-bandwidth calculation.  Its purpose is instead to identify which closed tunneling paths generate the local and non-local Josephson terms, which paths renormalize the singlet--triplet exchange, and why a single fourth-order many-body branch remains reciprocal even when its current--phase relation is shifted.

We use the gauge
\begin{align}
    \tau_{L1}&=t_{L1}e^{i\chi},
    &\tau_{R1}&=t_{R1}e^{-i\chi},\nonumber\\
    \tau_{L2}&=\mathcal P t_{L2},
    &\tau_{R2}&=t_{R2},
    \label{eq:app_gauge}
\end{align}
with
\begin{equation}
    2\chi=\frac{\pi\Phi}{\Phi_0},
    \qquad
    \varphi=\phi_L-\phi_R-2\chi,
    \qquad
    \Phi_0=\frac{h}{2e}.
    \label{eq:app_phases}
\end{equation}
We also define the two split-pair tunnel products
\begin{equation}
    T_L=t_{L1}t_{L2},
    \qquad
    T_R=t_{R1}t_{R2}.
    \label{eq:app_TLR}
\end{equation}

\subsection{Fourth-order low-energy Hamiltonian}
\label{app:fourth_order_formalism}

Let
\begin{equation}
    H=H_0+V,
    \qquad
    H_0=H_S+H_{\mathrm{DQD}},
    \qquad
    V=H_T,
\end{equation}
and let $\Pi$ project onto the low-energy manifold of a chosen DQD charge sector, with the superconducting reservoirs in their unperturbed ground state. The symbol $\Pi$ is used here to avoid confusion with the tunnel parity $\mathcal P$.  States inside the low-energy manifold are denoted by $|m\rangle$, whereas $|l\rangle$ denotes virtual states outside it.  For an unperturbed degenerate manifold of energy $E_0$ and $\Pi V\Pi=0$, the fourth-order Schrieffer--Wolff matrix can be written as
\begin{widetext}
\begin{align}
  \bigl(H_{\mathrm{eff}}^{(4)}\bigr)_{mm'}
  ={}&-
  \frac{1}{2}
  \sum_{l,l',m''}
  V_{ml}V_{lm''}V_{m''l'}V_{l'm'}
  \left[
  \frac{1}{(E_0-E_l)^2(E_0-E_{l'})}
  +
  \frac{1}{(E_0-E_l)(E_0-E_{l'})^2}
  \right]
  \nonumber\\
  &+
  \sum_{l,l',l''}
  \frac{
  V_{ml}V_{ll'}V_{l'l''}V_{l''m'}
  }{
  (E_0-E_l)(E_0-E_{l'})(E_0-E_{l''})
  } .
  \label{eq:app_SW_fourth}
\end{align}
\end{widetext}
This is the form used in the diagrammatic fourth-order treatment of the same parallel-DQD geometry in Ref.~\cite{Probst2016}.  The first line of Eq.~\eqref{eq:app_SW_fourth} contains the folded, or reducible, contributions that return to the low-energy manifold after two tunnel events.  The second line contains genuine fourth-order sequences with three high-energy intermediate states.  Both terms are required for a Hermitian effective Hamiltonian and for avoiding double counting of iterated lower-order processes.

For later use, it is convenient to define the particle- and hole-like charge excitation energies of dot $i$ within the atomic $(1,1)$ valley
\begin{align}
    &\delta_i^-
    =E_i(0)-E_i(1)=-\epsilon_i,
    \\
    &\delta_i^+
    =E_i(2)-E_i(1)=\epsilon_i+U_i.
    \label{eq:app_charge_costs}
\end{align}
Both quantities are positive for $-U_i<\epsilon_i<0$.  In the ZBA, a virtual state with odd occupation of a superconducting site additionally costs an energy of order $\Delta_b$.  Hence the denominators entering Eq.~\eqref{eq:app_SW_fourth} contain combinations such as $\Delta_b+\delta_i^{\pm}$ and $2\Delta_b+\delta_i^{\pm}+\delta_j^{\pm}$, depending on the sequence of tunneling processes.

A compact definition of any process coefficient used below is
\begin{equation}
    K_{\gamma}^{\eta}
    =
    \sum_{\pi\in(\gamma,\eta)}
    s_{\pi}
    \frac{M_{\pi}^{\eta}}
    {\Delta E_{\pi,1}\Delta E_{\pi,2}\Delta E_{\pi,3}}
    +K_{\gamma,\mathrm{fold}}^{\eta}.
    \label{eq:app_coefficient_definition}
\end{equation}
Here $\gamma\in\{\mathrm J,\mathrm{CT}\}$ distinguishes Josephson and cotunneling paths, $\eta\in\{\mathrm{sc},\mathrm{se}\}$ distinguishes spin-conserving and spin-exchange contributions, $\pi$ labels the ordering of the tunneling processes, $s_\pi=\pm1$ is its fermionic sign, $M_{\pi}^{\eta}$ contains the BCS coherence factors and spin matrix elements, and $\Delta E_{\pi,j}>0$ is the excitation energy of the $j$th virtual state.  Equation~\eqref{eq:app_coefficient_definition} keeps the functional dependence transparent without displaying lengthy
branch-dependent rational functions.

Representative particlelike and holelike split-pair charge sequences are
\begin{align}
 (1,1)&\rightarrow(2,1)\rightarrow(2,2)
 \rightarrow(1,2)\rightarrow(1,1),\nonumber\\
 (1,1)&\rightarrow(0,1)\rightarrow(0,0)
 \rightarrow(1,0)\rightarrow(1,1),
 \label{eq:app_representative_sequences}
\end{align}
respectively.  Mixed particle--hole sequences are also possible. Their fermionic exchange signs differ from those of purely particle-like sequences and are responsible for the fact that non-local cotunneling can favor either the singlet or the triplet.

\subsection{Gauge-invariant closed tunnel products}
\label{app:gauge_products}

The phase structure can be obtained without choosing a gauge when the tunnel events form a loop.  For split-pair transfer, local pair transfer, and non-local cotunneling, respectively, we define
\begin{align}
    \mathcal A_\nu
    &=e^{-i\phi_\nu}\tau_{\nu1}\tau_{\nu2},
    \label{eq:app_A_general}\\
    \mathcal C_{\nu i}
    &=e^{-i\phi_\nu}\tau_{\nu i}^{\,2},
    \label{eq:app_C_general}\\
    \mathcal B_\nu
    &=\tau_{\nu1}\tau_{\nu2}^{*}.
    \label{eq:app_B_general}
\end{align}
The combinations $\mathcal A_L\mathcal A_R^*$, $\mathcal C_{Li}\mathcal C_{Ri}^*$, and $\mathcal B_L\mathcal B_R^*$ are invariant under arbitrary local rephasings of
the dot and lead fermions.  Their real parts therefore determine the physical orbital-phase factors.

Using Eqs.~\eqref{eq:app_gauge}--\eqref{eq:app_TLR}, the split-pair amplitudes are
\begin{equation}
    \mathcal A_L
    =\mathcal P T_L e^{-i(\phi_L-\chi)},
    \qquad
    \mathcal A_R
    =T_R e^{-i(\phi_R+\chi)},
\end{equation}
which gives
\begin{equation}
    2\operatorname{Re}\left(\mathcal A_L\mathcal A_R^*\right)
    =2\mathcal P T_LT_R\cos\varphi.
    \label{eq:app_non-local_J_phase}
\end{equation}
Thus, at a fixed gauge-invariant phase $\varphi$, the non-local Josephson interference has no additional orbital-flux dependence.

For non-local cotunneling
\begin{equation}
    \mathcal B_L=\mathcal P T_Le^{i\chi},
    \qquad
    \mathcal B_R=T_Re^{-i\chi},
\end{equation}
and consequently
\begin{equation}
    2\operatorname{Re}\left(\mathcal B_L\mathcal B_R^*\right)
%    =2\mathcal P T_LT_R\cos(2\chi)
    =2\mathcal P T_LT_R
      \cos\left(\frac{\pi\Phi}{\Phi_0}\right).
    \label{eq:app_non-local_CT_phase}
\end{equation}
This contribution is independent of $\varphi$ and therefore does not carry a current directly, but it shifts the many-body energies and the positions of phase-dependent crossings.

For local pair transfer through dot 1
\begin{equation}
    2\operatorname{Re}\left(\mathcal C_{L1}\mathcal C_{R1}^*\right)
    =2t_{L1}^2t_{R1}^2\cos(\varphi-2\chi),
    \label{eq:app_local_J1_phase}
\end{equation}
whereas for dot 2 the tunnel-parity sign appears twice and cancels,
\begin{equation}
    2\operatorname{Re}\left(\mathcal C_{L2}\mathcal C_{R2}^*\right)
    =2t_{L2}^2t_{R2}^2\cos(\varphi+2\chi).
    \label{eq:app_local_J2_phase}
\end{equation}
Finally, a closed local cotunneling path contains products such as
\begin{equation}
    (\tau_{Li}\tau_{Ri}^{*})(\tau_{Ri}\tau_{Li}^{*})
    =|\tau_{Li}\tau_{Ri}|^2
    =t_{Li}^2t_{Ri}^2,
    \label{eq:app_local_CT_phase}
\end{equation}
so its phase, flux, and tunnel-parity factors cancel at this order.

The resulting phase structure is summarized in Table~\ref{tab:app_processes1}.
\begin{table*}[t]
\caption{Cross-lead fourth-order interference factors.  The coefficients multiplying the listed processes contain the virtual-state denominators, coherence factors and fermionic exchange signs. ``Direct current'' indicates whether the term contributes to $\partial E/\partial\phi$ at fixed flux.}
\label{tab:app_processes1}
\begin{ruledtabular}
\begin{tabular}{lcccc}
Process & Tunnel product & Phase factor & Spin structure & Direct current \\
\hline
Local Josephson through dot 1
& $t_{L1}^2t_{R1}^2$
& $\cos(\varphi-2\chi)$
& spin conserving
& yes \\
Local Josephson through dot 2
& $t_{L2}^2t_{R2}^2$
& $\cos(\varphi+2\chi)$
& spin conserving
& yes \\
Non-local Josephson
& $\mathcal P T_LT_R$
& $\cos\varphi$
& spin conserving and spin exchange
& yes \\
Non-local cotunneling
& $\mathcal P T_LT_R$
& $\cos(2\chi)$
& spin conserving and spin exchange
& no \\
Local cotunneling
& $t_{Li}^2t_{Ri}^2$
& constant
& spin conserving
& no
\end{tabular}
\end{ruledtabular}
\end{table*}

\subsection{Spin-conserving and spin-exchange decomposition}
\label{app:spin_decomposition}

In the $(1,1)$ sector, the spin-rotation-invariant Hamiltonian is diagonal in
the singlet and triplet subspaces.  Their projectors are
\begin{equation}
    \widehat P_S
    =\frac{1}{4}-\textbf{S}_1\,\cdot\,\textbf{S}_2,
    \qquad
    \widehat P_T
    =\frac{3}{4}+\textbf{S}_1\,\cdot\,\textbf{S}_2.
    \label{eq:app_spin_projectors}
\end{equation}
It is useful to introduce
\begin{equation}
    \widehat X
    =\widehat P_T-\widehat P_S
    =2\textbf{S}_1\,\cdot\,\textbf{S}_2+\frac{1}{2},
    \label{eq:app_exchange_operator}
\end{equation}
which satisfies
$\widehat X|S\rangle=-|S\rangle$ and
$\widehat X|T_m\rangle=+|T_m\rangle$. The effective Hamiltonian can then be
written as
\begin{equation}
    H_{\mathrm{eff}}^{(1,1)}(\varphi,\chi)
    =E_{\mathrm{av}}(\varphi,\chi)\,\mathbbm{1}
    +J_{\mathrm{ex}}(\varphi,\chi)\,\widehat X,
    \label{eq:app_Heff_ST}
\end{equation}
so that
\begin{equation}
    E_S=E_{\mathrm{av}}-J_{\mathrm{ex}},
    \qquad
    E_T=E_{\mathrm{av}}+J_{\mathrm{ex}},
    \qquad
    E_T-E_S=2J_{\mathrm{ex}}.
    \label{eq:app_EST}
\end{equation}
With this convention, $J_{\mathrm{ex}}>0$ favors the singlet and $J_{\mathrm{ex}}<0$ favors the triplet.

For identical superconducting reservoirs, the left- and right-lead sums factorize into the two non-local interference weights
\begin{align}
    \Lambda_{\mathrm J}(\varphi)
    &=T_L^2+T_R^2
      +2\mathcal P T_LT_R\cos\varphi
      =\left|T_R+\mathcal P T_Le^{i\varphi}\right|^2,
    \label{eq:app_LambdaJ}\\
    \Lambda_{\mathrm{CT}}(\chi)
    &=T_L^2+T_R^2
      +2\mathcal P T_LT_R\cos(2\chi)
      =\left|T_R+\mathcal P T_Le^{2i\chi}\right|^2.
    \label{eq:app_LambdaCT}
\end{align}
The complete fourth-order structure in the $(1,1)$ sector may consequently be organized as
\begin{align}
 E_{\mathrm{av}}={}&E_{11}^{(0)}+E_{\mathrm{const}}
 +K_{\mathrm J}^{\mathrm{sc}}\Lambda_{\mathrm J}(\varphi)
 +K_{\mathrm{CT}}^{\mathrm{sc}}\Lambda_{\mathrm{CT}}(\chi)
 \nonumber\\
 &+A_1t_{L1}^2t_{R1}^2\cos(\varphi-2\chi)
 +A_2t_{L2}^2t_{R2}^2\cos(\varphi+2\chi),
 \label{eq:app_Eav_full}\\
 J_{\mathrm{ex}}={}&
 K_{\mathrm J}^{\mathrm{se}}\Lambda_{\mathrm J}(\varphi)
 +K_{\mathrm{CT}}^{\mathrm{se}}\Lambda_{\mathrm{CT}}(\chi).
 \label{eq:app_Jex_full}
\end{align}
Here $E_{\mathrm{const}}$ contains the local cotunneling corrections and all other phase-independent spin-conserving terms. We use the abbreviations ``sc" and ``se" for spin-conserving and spin-exchange processes, respectively. The coefficients $A_i$, $K_{\gamma}^{\mathrm{sc}}$, and $K_{\gamma}^{\mathrm{se}}$ depend on $\epsilon_i$, $U_i$, $\Delta_b$, and the branch under consideration, but not on $\varphi$, $\chi$, or $\mathcal P$.  Folded terms from the first line of Eq.~\eqref{eq:app_SW_fourth} are included in these coefficients.

Expanding Eq.~\eqref{eq:app_Jex_full} yields the form used in the main text
\begin{equation}
    J_{\mathrm{ex}}(\varphi,\chi)
    =J_0
    +\mathcal P J_{\mathrm J}\cos\varphi
    +\mathcal P J_{\mathrm{CT}}\cos(2\chi),
    \label{eq:app_Jex_main_form}
\end{equation}
where, for the factorized form above
\begin{align}
    J_0
    &=\left(K_{\mathrm J}^{\mathrm{se}}
       +K_{\mathrm{CT}}^{\mathrm{se}}\right)
      \left(T_L^2+T_R^2\right),
    \label{eq:app_J0_coeff}\\
    J_{\mathrm J}
    &=2K_{\mathrm J}^{\mathrm{se}}T_LT_R,
    \qquad
    J_{\mathrm{CT}}
    =2K_{\mathrm{CT}}^{\mathrm{se}}T_LT_R.
    \label{eq:app_J_coeffs}
\end{align}
Any additional phase-independent exchange contribution may be absorbed into $J_0$. Eq.~\eqref{eq:app_Jex_main_form} is a decomposition of the same fourth-order corrections appearing in Eqs.~\eqref{eq:app_Eav_full} and \eqref{eq:app_Jex_full}; it is not an additional interaction to be counted a second time.

The Josephson spin-exchange diagrams are singlet favoring in the present spin-rotation-invariant model, corresponding to $K_{\mathrm J}^{\mathrm{se}}>0$ in the convention of
Eq.~\eqref{eq:app_EST}.  In contrast, the cotunneling diagrams contain both signs.  Purely electronlike exchange sequences favor the singlet, whereas mixed electron--hole sequences can favor the triplet.  A useful schematic estimate of their relative energy-denominator weights is
\begin{align}
    W_{\mathrm{ee}}
    &\sim
    \frac{1}{
    (\epsilon_{\mathrm{ch}}+\Delta_b)^2
    (2\epsilon_{\mathrm{ch}}+2\Delta_b)},
    \label{eq:app_weight_ee}\\
    W_{\mathrm{eh}}
    &\sim
    \frac{1}{
    (\epsilon_{\mathrm{ch}}+\Delta_b)^2
    (2\Delta_b)},
    \label{eq:app_weight_eh}
\end{align}
where $\epsilon_{\mathrm{ch}}$ denotes a typical DQD charge-excitation energy. These estimates are not substitutes for the full sums in Eq.~\eqref{eq:app_coefficient_definition}; they only show why triplet-favoring cotunneling becomes comparatively important when $\Delta_b$ is not much larger than the dot charging scales.

Eqs.~\eqref{eq:app_LambdaJ} and \eqref{eq:app_LambdaCT} also clarify the role of tunnel parity.  At zero flux, $\Lambda_{\mathrm{CT}}=(T_R+\mathcal PT_L)^2$.  For symmetric split-pair couplings, $T_L=T_R$, the non-local cotunneling interference is therefore constructive for $\mathcal P=+1$ and cancels for $\mathcal P=-1$.  At finite flux, the cancellation and enhancement conditions are displaced by the Aharonov--Bohm phase $2\chi$.

\subsection{Singlet--triplet crossing and current discontinuity}
\label{app:ST_crossing}

The singlet and triplet cross when
\begin{equation}
    J_{\mathrm{ex}}(\varphi_{\mathrm{ST}},\chi)=0.
\end{equation}
Using Eq.~\eqref{eq:app_Jex_main_form}, the crossing phase satisfies
\begin{equation}
    \cos\varphi_{\mathrm{ST}}
    =-
    \frac{
      J_0+\mathcal P J_{\mathrm{CT}}\cos(2\chi)
    }{
      \mathcal P J_{\mathrm J}
    }.
    \label{eq:app_ST_crossing_phase}
\end{equation}
A phase-dependent singlet--triplet transition exists only when
\begin{equation}
    \left|
      J_0+\mathcal P J_{\mathrm{CT}}\cos(2\chi)
    \right|
    \leq |J_{\mathrm J}|.
    \label{eq:app_ST_existence}
\end{equation}
The non-local cotunneling term therefore moves, creates, or removes the crossing even though it does not contribute directly to the current at fixed
flux.

The branch currents are
\begin{equation}
    I_{S/T}(\varphi)
    =\frac{2e}{\hbar}
      \frac{\partial E_{S/T}}{\partial\varphi},
\end{equation}
so their difference is
\begin{equation}
    I_T-I_S
    =\frac{4e}{\hbar}
      \frac{\partial J_{\mathrm{ex}}}{\partial\varphi}
    =-\frac{4e}{\hbar}
      \mathcal P J_{\mathrm J}\sin\varphi.
    \label{eq:app_current_jump}
\end{equation}
At an equilibrium singlet--triplet crossing, the ground-state energy is the lower envelope of the two branches.  If the slopes in Eq.~\eqref{eq:app_current_jump} differ, the zero-temperature current is discontinuous and its one-sided values are obtained from the branch selected on each side of the crossing.

\begin{figure}
    \centering
  %  \caption[]{$\mathcal{P}=1$, point A}
\includegraphics[width=\linewidth,height=0.3\textwidth]{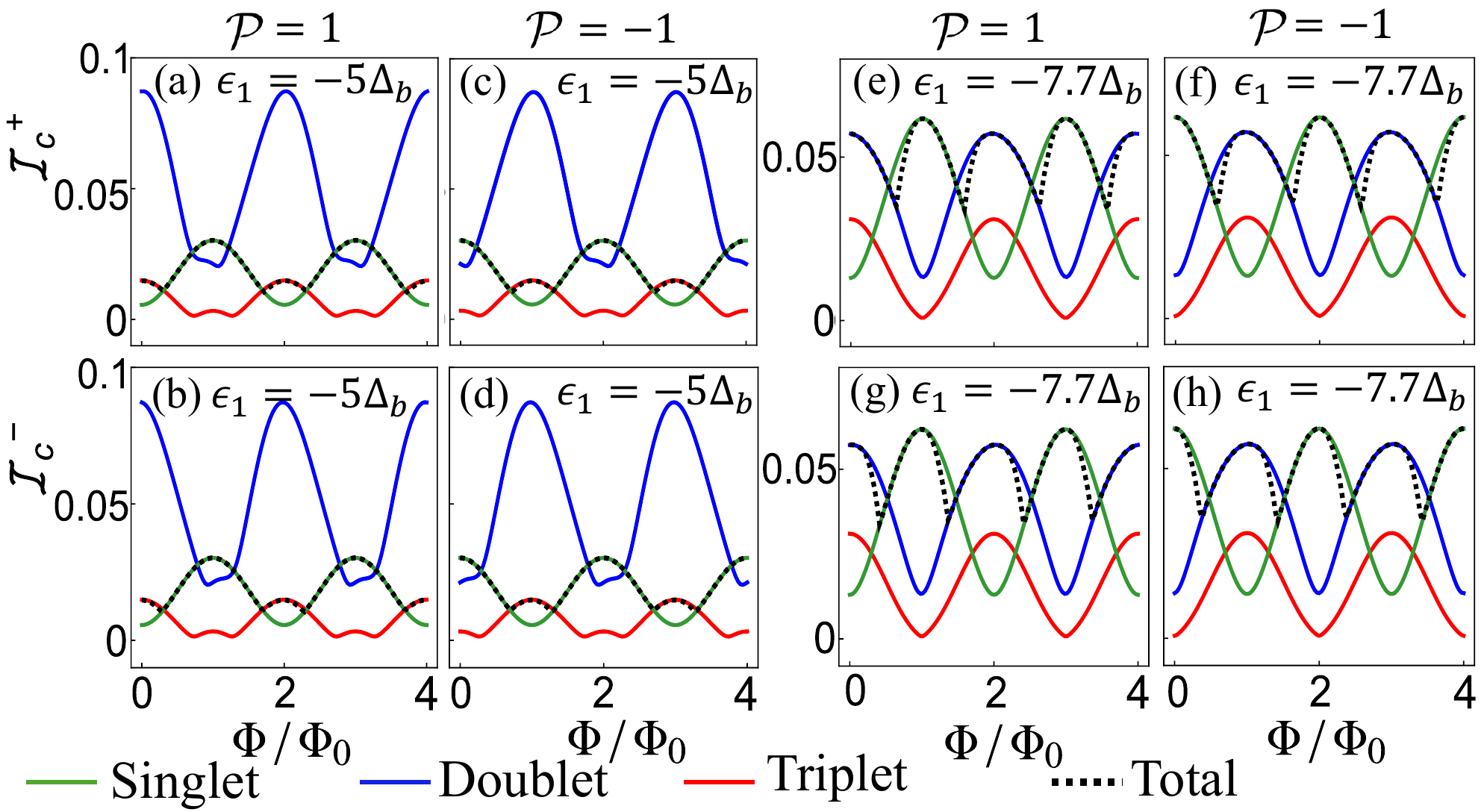}
\caption{Positive and negative dimensionless critical currents $\mathcal I_c^+$ and $\mathcal I_c^-$ given by different spin states as a function of magnetic flux $\Phi/\Phi_0$ for even $(\mathcal{P}=1)$ and odd $(\mathcal{P}=-1)$ tunnel parities at first-dot level energies $\epsilon_1=-5\Delta_b$ and $-7.7\Delta_b$. The total critical current given by the ground state energy follows different many body spin-state branches. We use $U=8\Delta_b$, $\epsilon_2=-1.5\Delta_b$, and $t_{L1}=\abs{t_{L2}}=t_{R1}=t_{R2}=0.5\Delta_b$.}
    \label{fig:7}
\end{figure}

\begin{figure}
    \centering
  %  \caption[]{$\mathcal{P}=1$, point A}
\includegraphics[width=\linewidth,height=0.4\textwidth]{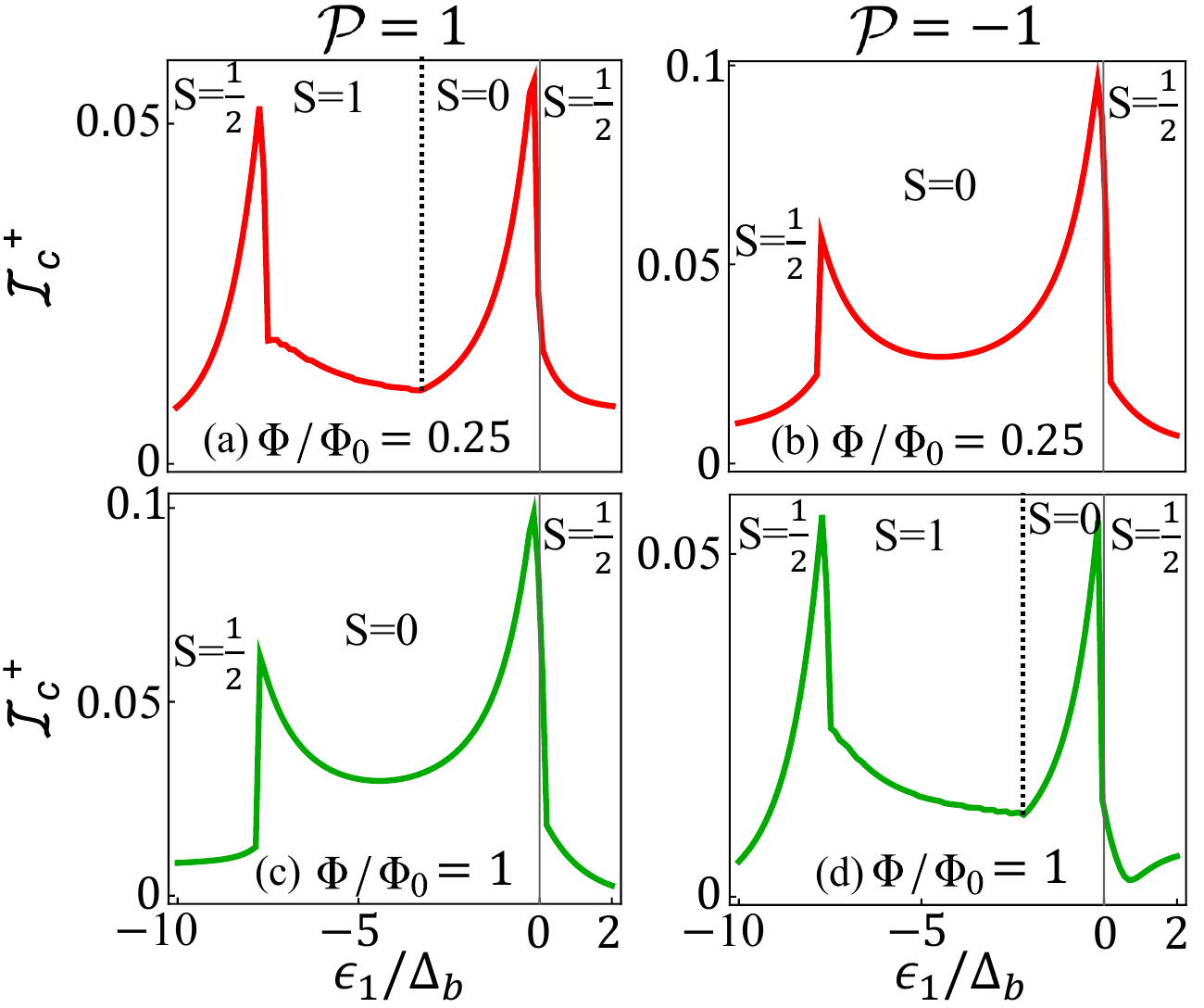}
\caption{Positive dimensionless critical current $\mathcal I_c^+$ as a function of the first-dot level $\epsilon_1$ for two orbital phases and both tunnel parities. The dominant total-spin character is indicated in the panels. We use $U=8\Delta_b$, $\epsilon_2=-1.5\Delta_b$, and $t_{L1}=\abs{t_{L2}}=t_{R1}=t_{R2}=0.5\Delta_b$.}
    \label{fig:8}
\end{figure}

%------------------------------------------------------
%\FloatBarrier
\section{Additional results on the parameter dependence of Josephson-current nonreciprocity}
\label{app:additional results}
%------------------------------------------------------
In this Appendix, we provide additional results illustrating the dependence of the Josephson-current nonreciprocity on key control parameters. We first present representative line cuts of the diode efficiency as a function of magnetic flux, followed by an analysis of the role of tunnel-coupling asymmetry and finite temperature. These results complement the main-text discussion by clarifying the robustness and tunability of the nonreciprocal response across different parameter regimes.
%-----------------------------------------------------
\begin{figure}
    \centering
  %  \caption[]{$\mathcal{P}=1$, point A}
\includegraphics[width=\linewidth,height=0.26\textwidth]{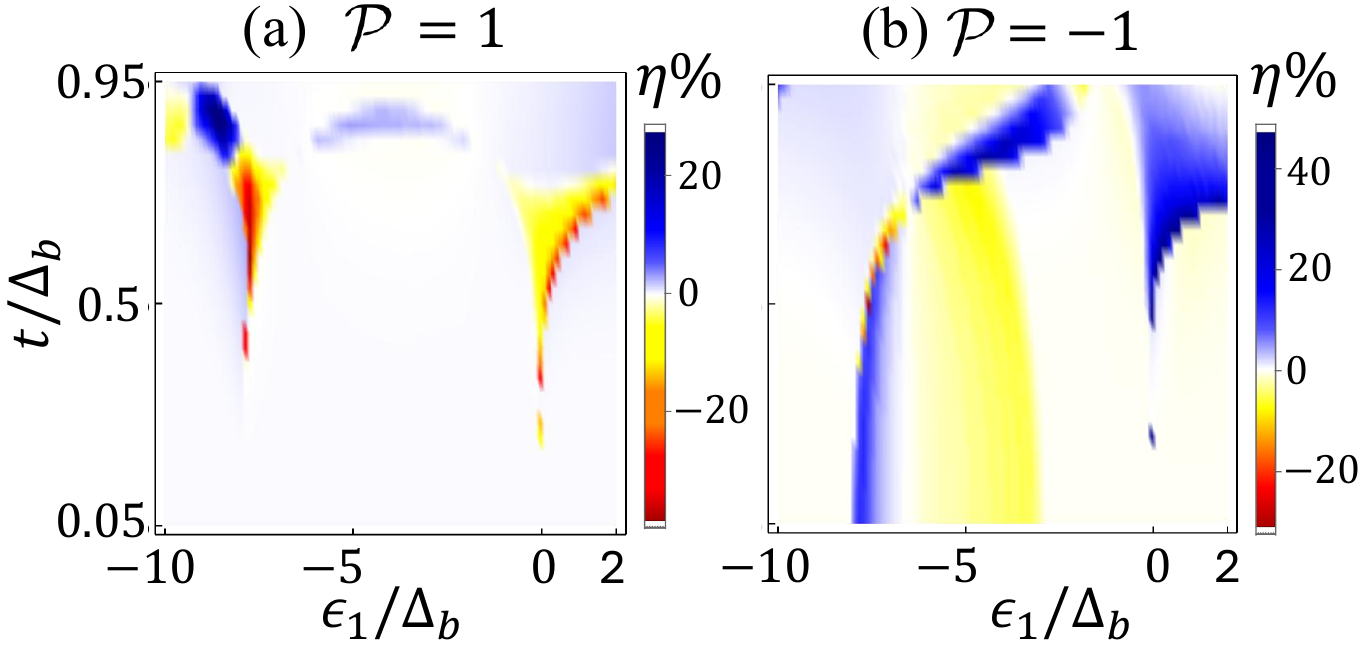}
\caption{Josephson-diode efficiency in the $(\epsilon_1, t)$ plane for (a) even and (b) odd total tunnel parity at fixed $\Phi/\Phi_0=0.76$. Here we define $t_{L1}=\abs{t_{L2}}=t_{R1}=t_{R2}=t$ and $U=8\Delta_b, \epsilon_2=-1.5\Delta_b$.}
    \label{fig:9}
\end{figure}
%----------------------------------------
\subsection{Line cuts through the flux-controlled spectrum}
\label{app:additional-flux-cuts}
%------------------------------------------------------
The two-dimensional maps in the main text emphasize the diode efficiency. Here we provide complementary one-dimensional line cuts in Fig.~\ref{fig:7} that identify ground state critical current mediated charge- and spin-sector transitions underlying their sharp structures.
%------------------------------------------
\begin{figure*}
    \centering
  %  \caption[]{$\mathcal{P}=1$, point A}
\includegraphics[width=0.9\linewidth,height=0.42\textwidth]{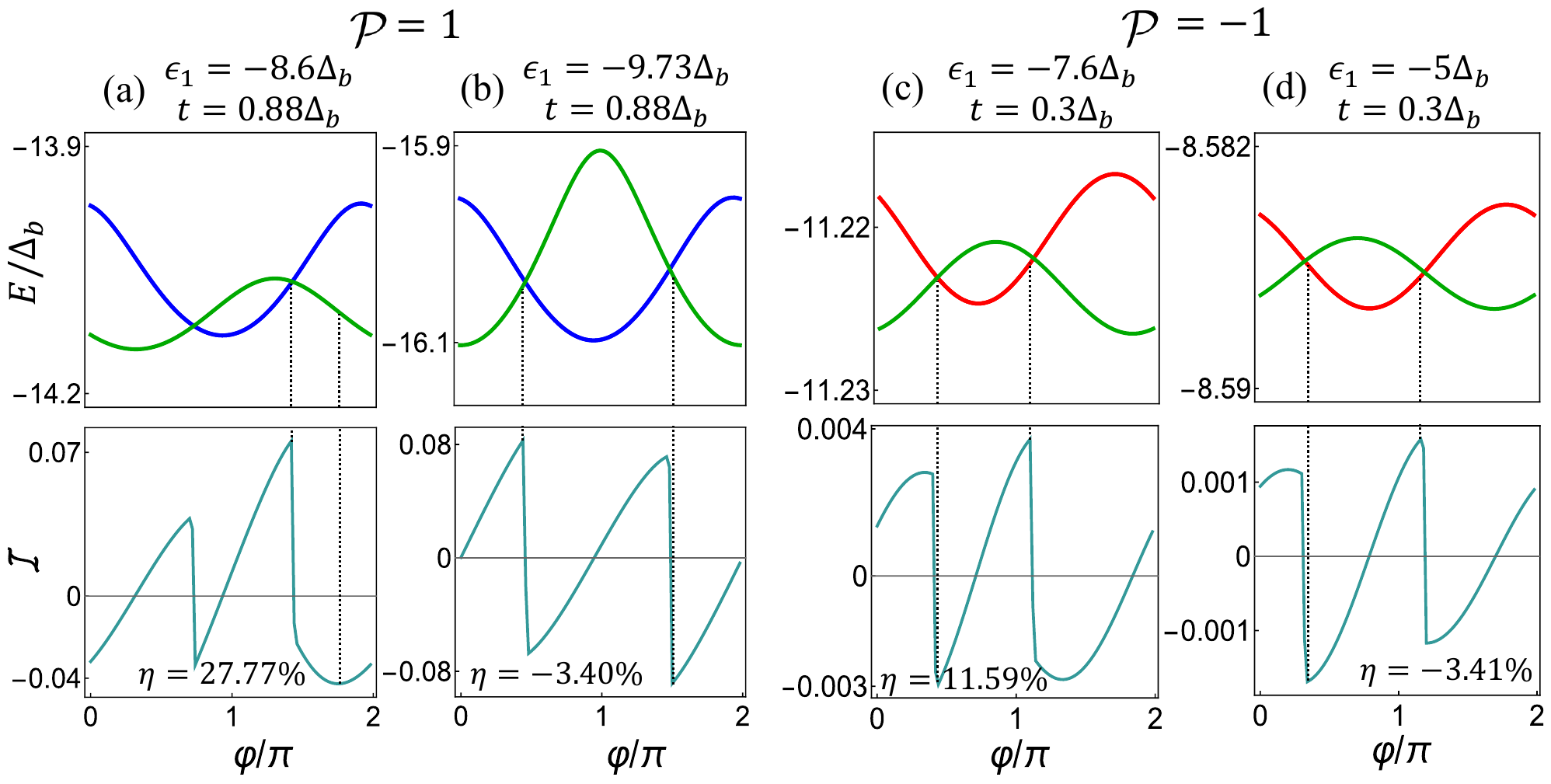}
\caption{Effect of variation in tunneling amplitude to the lowest phase-dependent many-body energies (upper row) and Josephson current (lower row) at different combination of parameters $(\mathcal{P},\epsilon_1,t)$ referring to several finite efficiency regimes in Fig.~\ref{fig:9}.}
    \label{fig:10}
\end{figure*}
%----------------------------------------------
%-----------------------------------------------
\begin{table*}[t]
\caption{Branch-selected efficiencies of Fig.~\ref{fig:10}}
\label{tab:app_processes2}
\begin{ruledtabular}
\begin{tabular}{lcccc}
Parity & Parameter & Singlet branch $\eta$ & Triplet branch $\eta$ & Doublet branch $\eta$  \\
\hline
even
& $\epsilon_1=-8.6\Delta_b, t=0.88\Delta_b$
& $0.87\%$
& $-1.56\%$
& $0.43\%$\\
even
& $\epsilon_1=-9.73\Delta_b, t=0.88\Delta_b$
& $0.809\%$
& $-1.32\%$
& $-1.51\%$\\
odd
& $\epsilon_1=-7.6\Delta_b, t=0.3\Delta_b$
& $0.015\%$
& $0.025\%$
& $0.34\%$\\
odd
& $\epsilon_1=-5\Delta_b, t=0.3\Delta_b$
& $0.004\%$
& $-0.0056\%$
& $-0.68\%$
\end{tabular}
\end{ruledtabular}
\end{table*}
%------------------------------------------
%\subsection{Critical-current cuts}
%\label{app:critical-current-cuts}
For $U=8\Delta_b$, the atomic charge-degeneracy points of dot 1 are $\epsilon_1=-8\Delta_b$ and $\epsilon_1=0$. The narrow peaks in Fig.~\ref{fig:8} occur close to these boundaries, where the ground-state envelope switches between the $(2,1)$, $(1,1)$, and $(0,1)$ sectors. Hybridization shifts the transition points slightly away from their atomic values. Additional kinks inside the singly occupied interval $-U<\epsilon_1<0$ mark changes between singlet- and triplet-dominated branches.

The parity dependence follows the non-local exchange factor $\mathcal P\cos(2\chi)$. At $\chi=\pi/8$, a triplet segment occurs for even tunnel parity, whereas at $\chi=\pi/2$ the corresponding segment occurs for odd tunnel parity. This complementarity is the one-dimensional counterpart of the flux-gate maps in Fig.~\ref{fig:3}. The curves in Fig.~\ref{fig:8} show only $\mathcal I_c^+$, by themselves they locate many-body transitions but do not quantify the diode response, which requires comparison with $\mathcal I_c^-$.

%-------------------------------------------
\begin{figure}
    \centering
  %  \caption[]{$\mathcal{P}=1$, point A}
\includegraphics[width=\linewidth,height=0.27\textwidth]{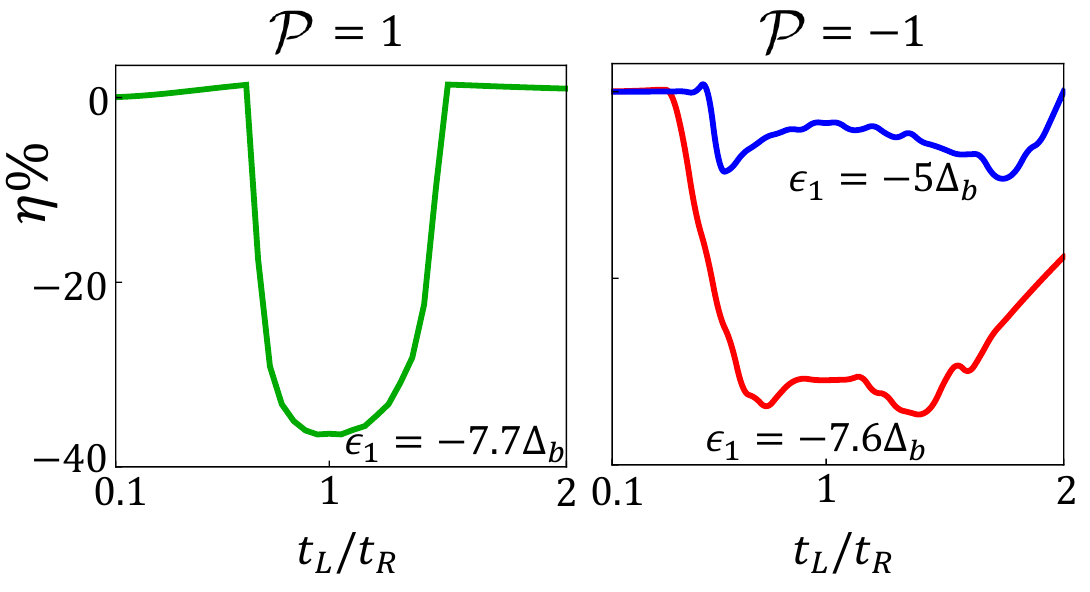}
\caption{Effect of asymmetric tunnel couplings of the QDs to the left and right leads: variation of Josephson-diode efficiency with the ratio of different tunneling parameters along left and right leads $t_L/t_R$, considering $t_{L1}=\abs{t_{L2}}=t_L$ and $t_{R1}=t_{R2}=t_R$ for (a) even and (b) odd total tunnel parity at fixed $\Phi/\Phi_0=0.76$. For $\mathcal{P}=1$, $\epsilon_1$ is chosen the same as in point A of Fig.~\ref{fig:5}. For $\mathcal{P}=-1$, the other parameters are the same as point C(red) and D(d) in Fig.~\ref{fig:5}.}
    \label{fig:11}
\end{figure}
%------------------------------------------------------
%----------------------------------------------
\begin{figure*}
    \centering
  %  \caption[]{$\mathcal{P}=1$, point A}
\includegraphics[width=0.9\linewidth,height=0.38\textwidth]{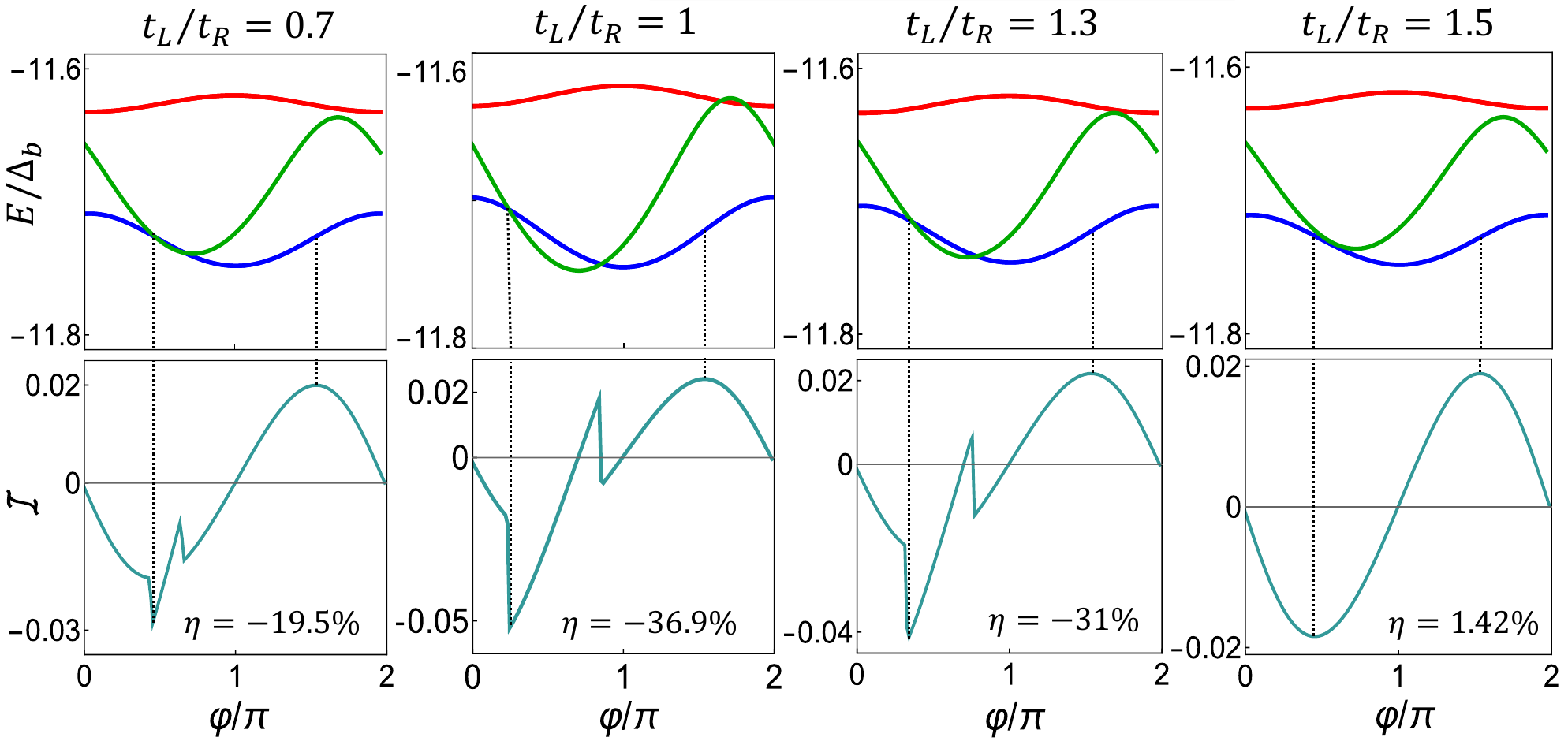}
\caption{Effect of asymmetric tunnel coupling to the lowest phase-dependent many-body energies (upper row) and Josephson current (lower row) at $t_L/t_R$. The color codes for different branches maintains the same as in Fig.~\ref{fig:5}. Here we define $t_{L1}=t_{L2}=t_{L}$ and $t_{R1}=t_{R2}=t_{R}$ and the parameters are as follows: $U=8\Delta_b$, $\epsilon_1=-7.786\Delta_b, \epsilon_2=-1.5\Delta_b, \Phi/\Phi_{0}=0.76, \mathcal{P}=1$.}
    \label{fig:12}
\end{figure*}
%------------------------------------------------------
%----------------------------------------------
\begin{table*}[t]
\caption{Branch-selected efficiencies of Fig.~\ref{fig:12}}
\label{tab:app_processes3}
\begin{ruledtabular}
\begin{tabular}{lcccc}
Parity & Parameter & Singlet branch $\eta$ & Triplet branch $\eta$ & Doublet branch $\eta$  \\
\hline
even
& $t_L/t_R=0.7$
& $-0.33\%$
& $-0.105\%$
& $1.48\%$\\
even
& $t_L/t_R=1$
& $-0.373\%$
& $-0.12\%$
& $1.78\%$\\
even
& $t_L/t_R=1.3$
& $-0.35\%$
& $-0.111\%$
& $1.615\%$\\
even
& $t_L/t_R=1.5$
& $-0.324\%$
& $-0.101\%$
& $1.42\%$
\end{tabular}
\end{ruledtabular}
\end{table*}
%------------------------------------------
\subsection{Effect of symmetric tunnel coupling on nonreciprocity in supercurrent}\label{app:tunnel}
%------------------------------------------------------
Fig.~\ref{fig:9} shows the diode efficiency in the $(\epsilon_1,t)$ plane for $t_{L1}=|t_{L2}|=t_{R1}=t_{R2}=t$. Increasing $t$ strengthens both the local Josephson processes and the non-local pair-transfer and exchange processes, thereby modifying the hybridization and the relative dispersion of the competing many-body branches. This distinction is particularly important near the charge-sector boundaries, where the doublet branches are predominantly governed by local processes, while the competing singlet or triplet branches in the $(1,1)$ sector contain substantial non-local contributions. Their different phase dependences shift the branch crossings and modify the two critical-current extrema unequally as $t$ is varied.
%----------------------------------------------------------------
\begin{figure*}
    \centering \includegraphics[width=\linewidth,height=0.28\textwidth]{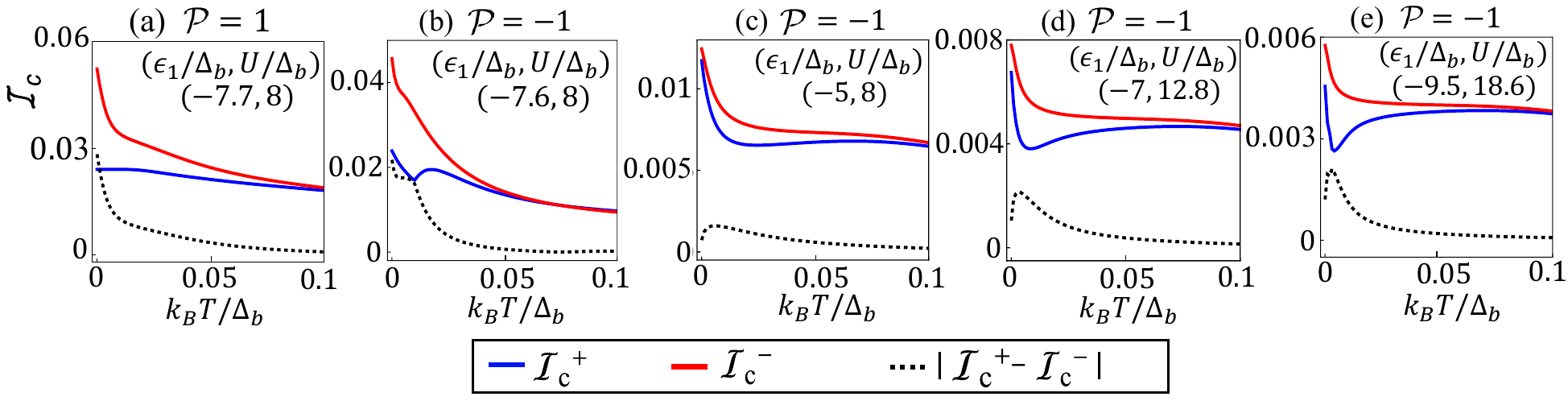}
    \caption{%\textcolor{red}{Check and rewrite the sentence: }
    Temperature dependence in the forward and reverse critical currents and diode response through the different in critical currents $\Delta\mathcal{I}=\abs{\mathcal{I}_c^+-\mathcal{I}_c^-}$ at different parameter regimes of $(\mathcal{P},\epsilon_1,U)$, following Fig.~\ref{fig:6}. All other parameters are the same as in Fig.~\ref{fig:6}.}
    \label{fig:13}
\end{figure*}

For $\mathcal P=1$, the pronounced finite-efficiency region near $\epsilon_1\simeq-U$ originates from singlet--doublet competition. As illustrated in Fig.~\ref{fig:10}(a,b), varying $(\epsilon_1,t)$ shifts the crossing relative to the critical phase differences and can change which branch determines $\mathcal{I}_c^{+}$ and $\mathcal{I}_c^{-}$, resulting in a strong variation and even a sign reversal of $\eta$. A similar mechanism occurs for $\mathcal P=-1$, where triplet--doublet and singlet--triplet rearrangements control the response (see Fig.~\ref{fig:10}(c,d)). Thus, the strongly nonmonotonic tunnel-coupling dependence reflects the competition between local and non-local transport processes through their modification of the phase-dependent many-body branch structure, rather than a simple increase of the supercurrent with $t$.

%------------------------------------------------------
\subsection{Effect of tunnel-coupling asymmetry}\label{app:tunnel_asy}
%------------------------------------------------------
We also extend the study of tunneling parameter dependence on current nonreciprocity by considering same coupling of both QDs to left (right) leads as $t_{L1}=t_{L2}=t_L (t_{R1}=t_{R2}=t_R)$ and introduce an asymmetry parameter $t_L/t_R$ along the left and right junctions, leading to enhanced inversion asymmetry at non-zero $t_L/t_R$. 

Fig.~\ref{fig:11} shows the variation of the diode efficiency with the left--right tunnel-coupling ratio $t_L/t_R$. The response is generally nonmonotonic, indicating that increasing tunnel asymmetry does not necessarily enhance rectification. Microscopically, the local Josephson contributions scale as $t_L^2t_R^2$, whereas the non-local Josephson and cotunneling terms contain the interference factors as shown in Eqs.~\ref{eq:app_LambdaJ} and \ref{eq:app_LambdaJ}. Thus, varying $t_L/t_R$ changes the interference weights between local and non-local processes and simultaneously modifies the exchange splitting and the position of the competing many-body branch crossings.

The importance of this branch rearrangement is illustrated in Fig.~\ref{fig:12}. For $\mathcal{P}=1$, the efficiency changes from $\eta=-19.5\%$ at $t_L/t_R=0.7$ to $-36.9\%$ at the symmetric point, followed by $-31\%$ at $t_L/t_R=1.3$ and a nearly reciprocal response, $\eta=1.42\%$, at $t_L/t_R=1.5$. This evolution reflects the displacement of the singlet--doublet crossing relative to the phases at which $\mathcal{I}_c^{+}$ and $\mathcal{I}_c^{-}$ are reached. For $\mathcal{P}=-1$, the reversed sign of the non-local interference further modifies the singlet--triplet or triplet--doublet ordering, resulting in the pronounced nonmonotonic behavior of $\eta(t_L/t_R)$. Hence, large rectification is obtained when tunnel asymmetry positions a many-body branch rearrangement close to one critical-current extremum while leaving the opposite one comparatively unaffected.

%------------------------------------------------------
\subsection{Critical current variation with temperature along line-cut of nonreciprocity}\label{app:temp}

Fig.~\ref{fig:13} resolves the temperature evolution of the positive and negative critical currents underlying the efficiency curves shown in Fig.~\ref{fig:6}. For each representative parameter set, we plot $\mathcal{I}_c^{+}$, $\mathcal{I}_c^{-}$, and the absolute asymmetry $\Delta \mathcal{I}_c=|\mathcal{I}_c^{+}-\mathcal{I}_c^{-}|$. Near the charge-sector crossovers, thermal occupation progressively weakens the branch-selection mechanism responsible for the large zero-temperature rectification. For even tunnel parity, $\mathcal{I}_c^{-}$ is suppressed more rapidly than $\mathcal{I}_c^{+}$, so that both $\Delta \mathcal{I}_c$ and $|\eta|$ decrease monotonically with temperature. For the odd-parity, triplet--doublet crossover, the weak low-temperature nonmonotonicity of $\eta$ does not correspond to an enhancement of the absolute diode response, since $\Delta I_c$ already decreases in the same temperature range. 

A qualitatively different behavior appears inside the triplet-dominated $(1,1)$ sector. There, a weak finite temperature can suppress one critical-current direction more strongly than the other because the excitation gap and the current carried by the nearby singlet-like state are different at the two critical phase differences. Consequently, $\Delta I_c$ develops a finite-temperature maximum for suitable parameters, demonstrating a genuine thermally assisted enhancement of the nonreciprocal current. At higher temperatures, additional many-body states become populated and the distinction between the two current directions is progressively averaged out, driving both $\Delta \mathcal{I}_c$ and $\eta$ toward zero. Fig.~\ref{fig:13} therefore confirms that the nonmonotonic temperature dependence of the normalized efficiency in Fig.~\ref{fig:6} can have two distinct origins: a reduction of the denominator $\mathcal{I}_c^{+}+\mathcal{I}_c^{-}$ near a charge-sector crossover, or a genuine enhancement of the absolute critical-current asymmetry inside the triplet-dominated regime. 
%------------------------------------------------------
\section{Zeeman and Spin-orbit coupling}\label{app:Zeeman_Rashba}
%------------------------------------------------------
\begin{figure}
    \centering
  %  \caption[]{$\mathcal{P}=1$, point A}
\includegraphics[width=\linewidth,height=0.26\textwidth]{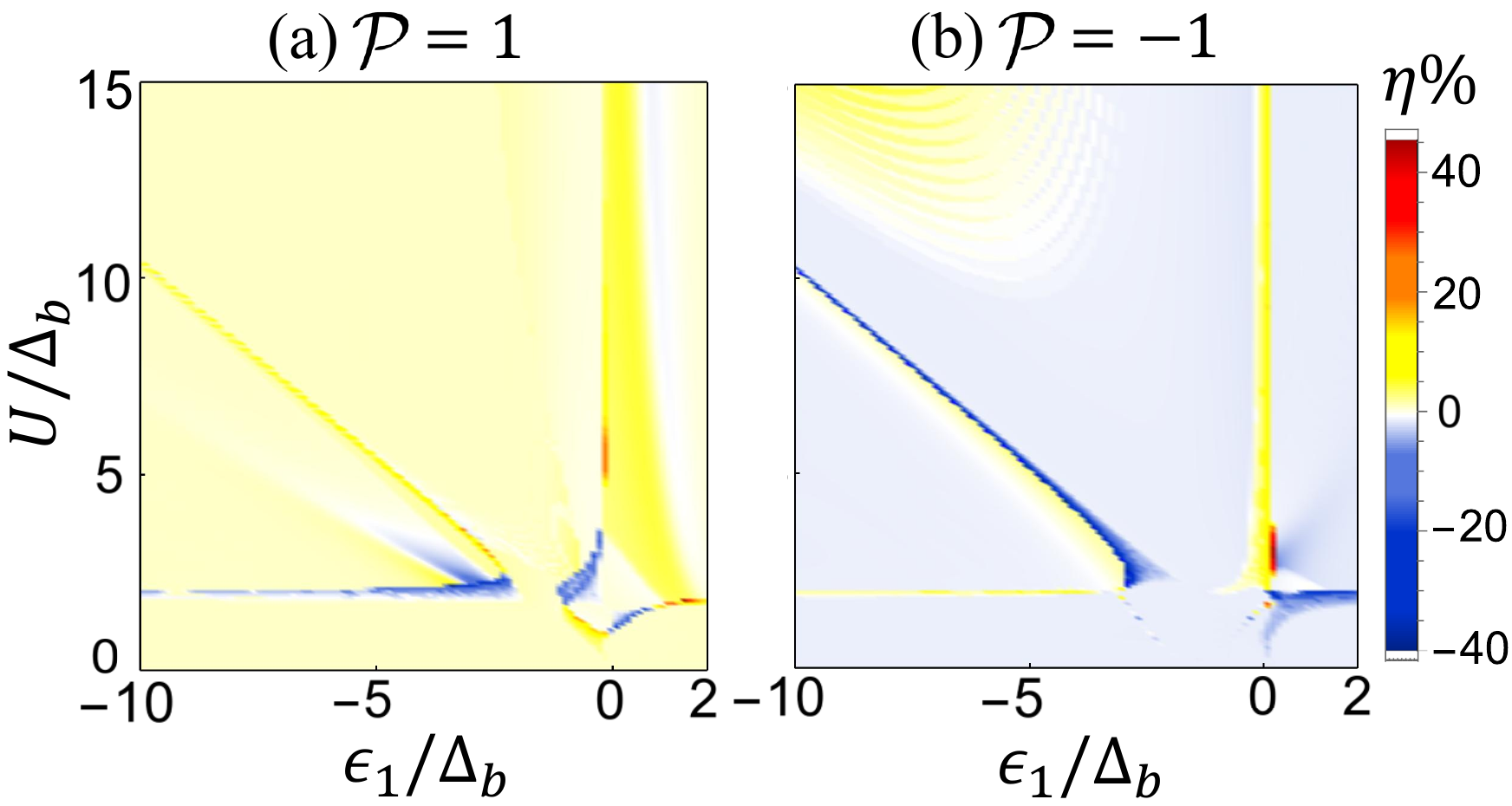}
\caption{Josephson-diode efficiency in the $(\epsilon_1,U)$ plane at zero orbital flux for (a) even and (b) odd tunnel parity. We use $\epsilon_2=-1.5\Delta_b$, $B_z=0.01\Delta_b$, $\alpha_{L1}=0.15\Delta_b$, $\alpha_{\nu i}=0$ on the other links, and $t_{L1}=\abs{t_{L2}}=t_{R1}=t_{R2}=0.5\Delta_b$.}
    \label{fig:14}
\end{figure}

The Zeeman term acts only on the dots
\begin{equation}
    H_Z=B_z\sum_{i=1,2}
    \left(\hat n_{i\uparrow}-\hat n_{i\downarrow}\right).
    \label{eq:H_z}
\end{equation}
With this convention, the two spin levels on each dot are shifted by $\pm B_z$ and their splitting is $2B_z$. The orbital flux $\Phi$ and the Zeeman energy $B_z$ are treated as independent control parameters.

Finally, we include the spin-orbit coupling (SOC) term through,
\begin{equation}
    H_{\mathrm{SO}}
    =\sum_{\nu\mathbf{k}j\sigma}
    \left(
        i s_{\sigma}\alpha_{\nu j}e^{ia_{\nu j}}
        d_{j\sigma}^{\dagger}c_{\mathbf{k}\nu\sigma}
        +\mathrm{h.c.}
    \right),
    \label{eq:H_RSOI}
\end{equation}
where $j=1,2$, $s_{\uparrow}=+1$ and $s_{\downarrow}=-1$. 
The orbital-flux and spin-orbit coupling mechanisms are analyzed separately. In the zero-flux calculations of Sec.~\ref{app:Zeeman_Rashba}, we set $\Phi/\Phi_0=0$ and hence  Peierls factor $a_{\nu j}=0$. 
If orbital flux and spin-orbit coupling were included simultaneously, the latter would carry the same orbital Peierls factors as the scalar tunneling amplitude on each link. Moreover, in order to preserve TRS, the SOC amplitudes $\alpha_{\nu j}$ have to be real. The SOC generates opposite tunnel phases for the two spin projections. It is the $S_z$-conserving component of a spin-orbit-active tunnel barrier: it can split or mix states within a fixed $S_z$ sector, but it does not produce spin flips in the chosen basis. An asymmetric distribution of $\alpha_{\nu j}$ breaks the inversion symmetry of the junction. In the calculations below, the spin-orbit coupling term is applied only to the $L1$ link, $\alpha_{L1}\neq0$, with all other $\alpha_{\nu j}=0$.

\subsection{Gate dependence of the diode response}
\label{app:gate_zeeman_soc}

Figure~\ref{fig:14} shows the diode efficiency in the $(\epsilon_1,U)$ plane at finite $B_z$ and $\alpha_{L1}$.
The strongest nonreciprocal response is concentrated near the boundaries where the dominant many-body ground state changes.
In the atomic limit, the charge-sector boundaries associated with dot~1 occur near $\epsilon_1=0$ and $\epsilon_1=-U$, while the fixed value $\epsilon_2=-1.5\Delta_b$ introduces an additional boundary at $U=1.5\Delta_b$. Finite hybridization shifts and broadens these transition lines. Away from these boundaries, the ground-state branch is only weakly reconstructed and the junction remains nearly reciprocal.
By contrast, close to charge- and spin-sector crossings, the Zeeman and spin-orbit terms cause the positive and negative critical currents to probe inequivalent portions of the strongly anharmonic many-body ground-state envelope, producing sizable rectification.

Changing the tunnel parity $\mathcal P$ modifies the location and sign of these regions through the non-local interference processes. Figure~\ref{fig:14} therefore demonstrates that symmetry breaking by $B_z$ and $\alpha_{L1}$ enables the diode effect, whereas large efficiencies additionally require proximity to a many-body branch rearrangement.

\subsection{Combined effect of Zeeman field and asymmetric spin-orbit tunneling}
\label{app:zeeman_soc_spectrum}

\begin{figure}
    \centering
\includegraphics[width=\linewidth,height=0.36\textwidth]{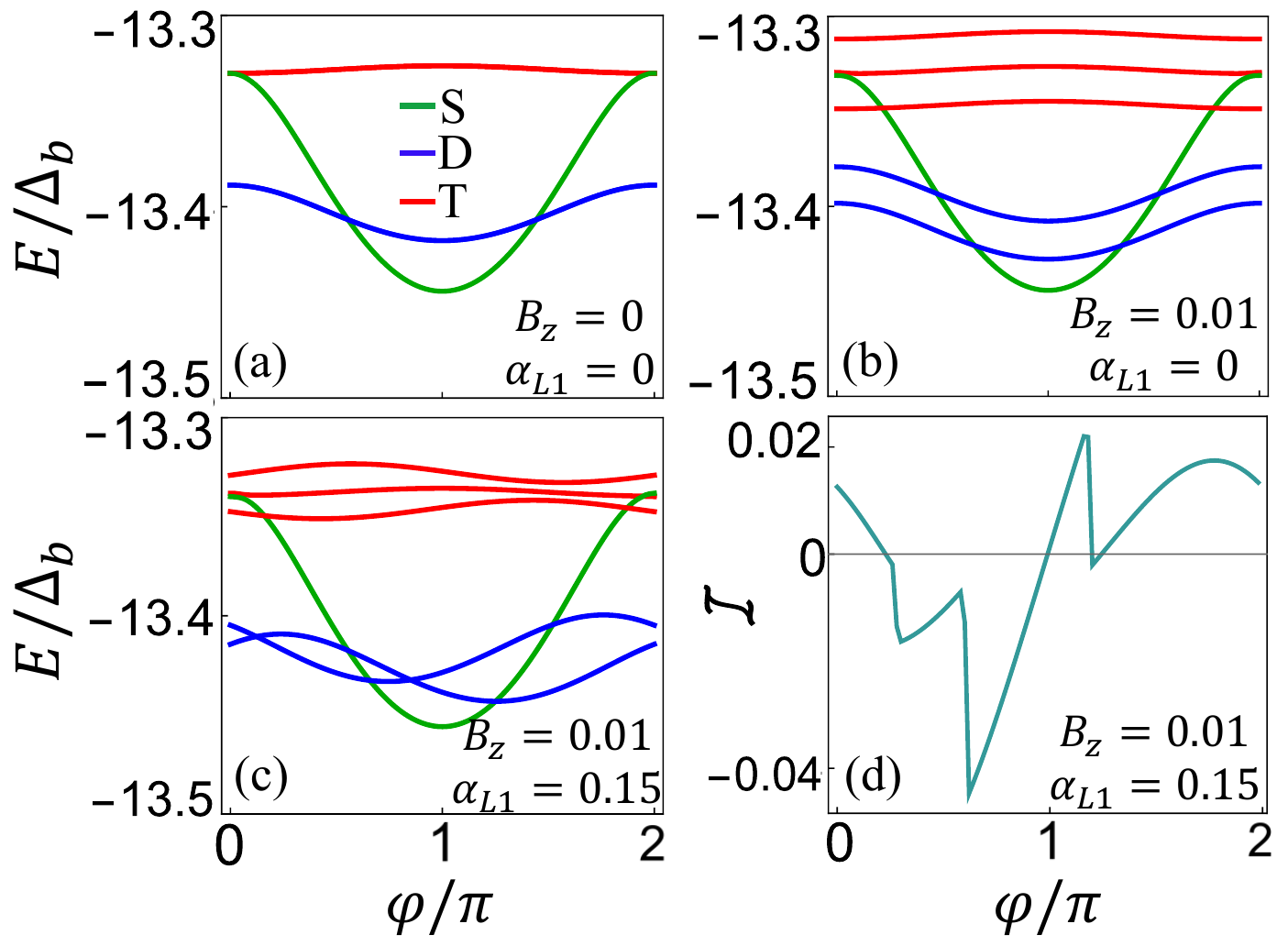}
\caption{Evolution of the lowest many-body energies and Josephson current at odd tunnel parity. (a) In the absence of Zeeman and spin-orbit terms, singlet- and doublet-like branches form the ground-state envelope. (b) A Zeeman field splits the spin multiplets while preserving the phase symmetry of the spectrum. (c) Asymmetric spin-orbit tunneling distorts the spectrum when combined with the Zeeman field, producing the nonreciprocal current in (d). At finite $\alpha_{L1}$, spin labels refer to the dominant character obtained by continuity from $\alpha_{L1}=0$. The parameters are $\epsilon_1=-9.5\Delta_b$, $\epsilon_2=-1.5\Delta_b$, and $U=9.75\Delta_b$.
}
    \label{fig:15}
\end{figure}

The microscopic origin of the zero-flux diode response is illustrated in Fig.~\ref{fig:15}. In the absence of both perturbations, the phase-dependent spectrum retains the symmetry required for reciprocal critical currents. A finite Zeeman field splits the doublet and triplet multiplets and shifts their crossings with the singlet branch. However, for $\alpha_{L1}=0$, the ground-state spectrum remains symmetric with respect to the superconducting phase, and therefore
$\mathcal{I}_c^+=\mathcal{I}_c^-$.

\begin{figure}
    \centering
\includegraphics[width=\linewidth,height=0.23\textwidth]{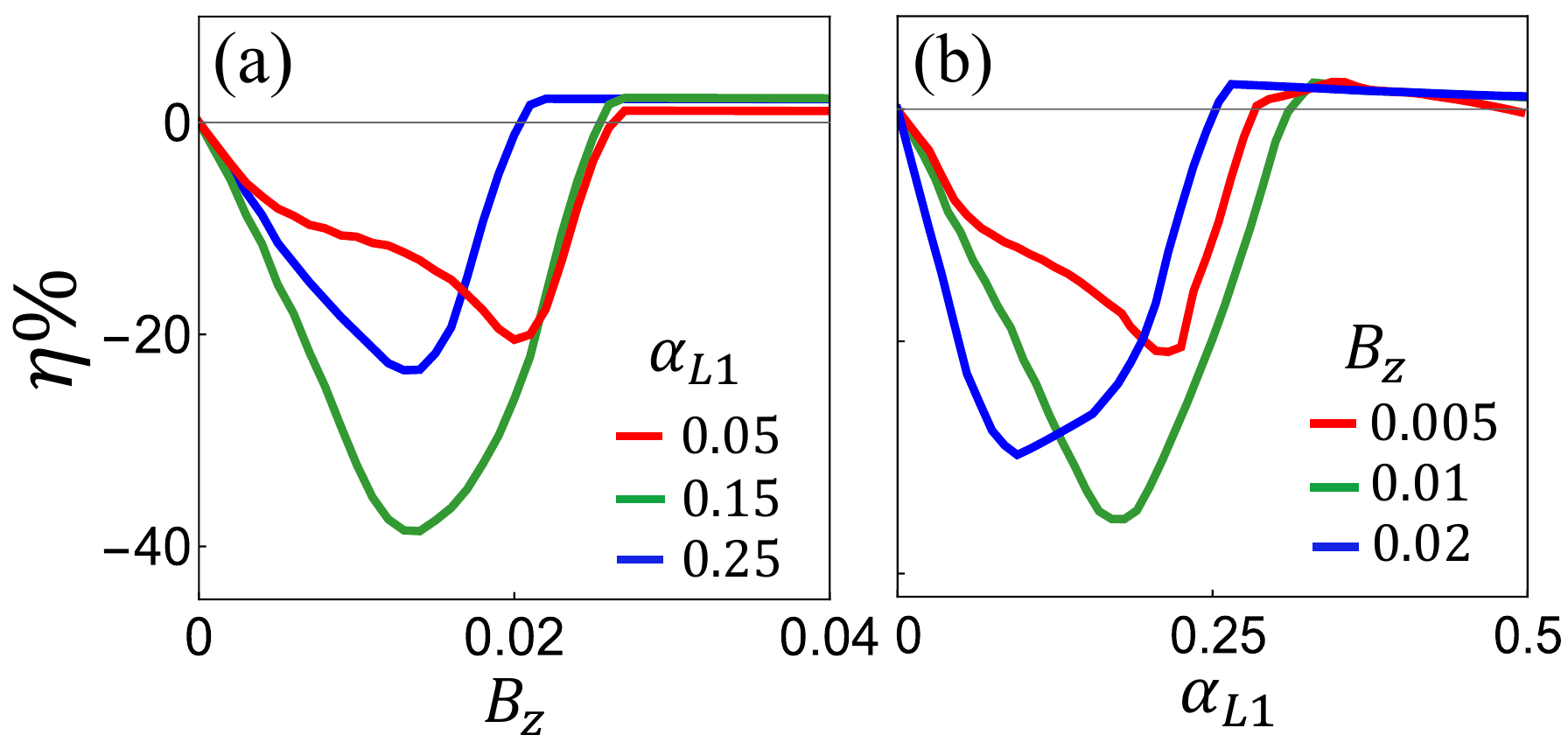}
\caption{Diode efficiency as a function of (a) Zeeman energy for several spin-orbit tunneling amplitudes and (b) spin-orbit tunneling amplitude for several Zeeman energies. The remaining parameters are $\epsilon_1=-9.5\Delta_b$, $\epsilon_2=-1.5\Delta_b$, $U=9.75\Delta_b$ and odd total tunnel parity.}
    \label{fig:16}
\end{figure}

Conversely, spin-orbit tunneling alone preserves time-reversal symmetry and is likewise insufficient to produce rectification.

When $B_z\neq0$ and $\alpha_{L1}\neq0$ act simultaneously, the two spin projections acquire inequivalent tunneling phases in the presence of the Zeeman splitting. The resulting deformation of the phase-dependent many-body spectrum removes its reflection symmetry and shifts the competing branch crossings to inequivalent phases. Consequently, the positive and negative extrema of the current--phase relation are modified differently, yielding $\mathcal{I}_c^+\neq \mathcal{I}_c^-$, as shown in Fig.~\ref{fig:15}(d).
At finite spin-orbit coupling, the spin sectors are generally mixed. The singlet, doublet, and triplet labels therefore refer to the dominant adiabatic character obtained from $\alpha_{L1}=0$ limit.

The corresponding dependence on the symmetry-breaking parameters is shown in Fig.~\ref{fig:16}.
The efficiency is nonmonotonic as either $B_z$ or $\alpha_{L1}$ is varied and vanishes when either perturbation is absent. At intermediate couplings, the deformation of the many-body branch structure is strongest and efficiencies approaching $|\eta|\simeq40\%$ are obtained. For larger couplings, the relevant branch crossings are displaced away from the critical-current-phases and the rectification is reduced. Thus, similarly to the orbital-flux mechanism discussed in the main text, the magnitude of the diode response is controlled not only by symmetry breaking but also by the position and phase dispersion of the competing many-body branches.

The orbital-flux and Zeeman-spin-orbit mechanisms are microscopically distinct, but they share the same organizing principle. A symmetry-breaking phase shift acts on a strongly anharmonic many-body current-phase relation, while Coulomb interactions and non-local exchange place singlet, doublet, and triplet branches close enough in energy to reshape the positive and negative current extrema differently. The tunnel parity then provides an additional discrete control of the non-local interference and of the spin sector that participates in the diode response.

\FloatBarrier

%\end{appendix}

\end{document}